\documentclass[a4paper,aps,prb,twocolumn,notitlepage,intlimits,superscriptaddress,8pt,floatfix]{revtex4-1}
\usepackage{amsmath,amsthm,amssymb,dsfont,mathrsfs,bbm,xfrac,xspace}
\usepackage[caption=false]{subfig}
\usepackage{mathtools}
\usepackage{multirow}
\usepackage[usenames,dvipsnames]{xcolor}
\usepackage[english]{babel}
\usepackage{hyperref}
\hypersetup{pdfauthor={Benetti Pietracaprina Parisi Sicuro},pdftitle={Soft modes},%
            colorlinks, linktocpage=true, pdfstartpage=3, pdfstartview=FitV,%
    breaklinks=true, pdfpagemode=UseNone, pageanchor=true, pdfpagemode=UseOutlines,%
    plainpages=false, bookmarksnumbered, bookmarksopen=true, bookmarksopenlevel=1,%
    hypertexnames=true, pdfhighlight=/O,%
    urlcolor=orange, linkcolor=blue, citecolor=red 
        }

\usepackage[T1]{fontenc}
\usepackage{lmodern}
\usepackage[cal=boondoxo]{mathalfa}

\newcommand{\bdelta}{{\boldsymbol\delta}}
\newcommand{\mG}{{\mathcal G}}

\newcommand{\bG}{{\boldsymbol{\mathsf G}}}
\newcommand{\bM}{{\boldsymbol{\mathsf M}}}
\newcommand{\bI}{{\boldsymbol{\mathsf I}}}

\newcommand{\bp}{{\boldsymbol{\mathsf p}}}
\newcommand{\bk}{{\boldsymbol{\mathsf k}}}
\newcommand{\bt}{{\boldsymbol{\mathsf t}}}
\DeclareMathOperator{\e}{e}
\DeclareMathOperator{\dd}{d}
\DeclareMathOperator{\tr}{Tr}
\DeclareMathOperator{\Imm}{Im}
\newcommand{\EE}[1]{\mathbb{E}\left[{#1}\right]}

\allowdisplaybreaks
\begin{document}
\title{Mean-field model for the density of states of jammed soft spheres}
\author{Fernanda P.C. Benetti}\email{fernanda.benetti@roma1.infn.it}
\affiliation{Dipartimento di Fisica, Sapienza Universit\`a di Roma, P.le A. Moro 2, I-00185, Rome, Italy}
\author{Giorgio Parisi}\email{giorgio.parisi@roma1.infn.it}
\affiliation{Dipartimento di Fisica, Sapienza Universit\`a di Roma, INFN -- Sezione di Roma1, and CNR-NANOTEC UOS Roma, P.le A. Moro 2, I-00185, Rome, Italy}
\author{Francesca Pietracaprina}\email{pietracaprina@irsamc.ups-tlse.fr}
\affiliation{Dipartimento di Fisica, Sapienza Universit\`a di Roma, P.le A. Moro 2, I-00185, Rome, Italy}
\affiliation{Laboratoire de Physique Th\'eorique, IRSAMC, Universit\'e de Toulouse, CNRS, UPS, France}

\author{Gabriele Sicuro}\email{gabriele.sicuro@roma1.infn.it}
\affiliation{Dipartimento di Fisica, Sapienza Universit\`a di Roma, P.le A. Moro 2, I-00185, Rome, Italy}

\date{\today}
\begin{abstract}We propose a class of mean-field models for the isostatic transition of systems of soft spheres, in which the contact network is modeled as a random graph and each contact is associated to $d$ degrees of freedom. We study such models in the hypostatic, isostatic, and hyperstatic regimes. The density of states is evaluated by both the cavity method and exact diagonalization of the dynamical matrix. We show that the model correctly reproduces the main features of the density of states of real packings and, moreover, it predicts the presence of localized modes near the lower band edge. Finally, the behavior of the density of states $D(\omega)\sim\omega^\alpha$ for $\omega\to 0$ in the hyperstatic regime is studied. We find that the model predicts a nontrivial dependence of $\alpha$ on the details of the coordination distribution.
\end{abstract}
\maketitle
\section{Introduction}
While the vibrational behavior of crystalline solids --- the density of states and heat capacity, for example --- is well known, the disorder present in the structure of non-crystalline systems such as glasses, granular materials, and foams leads to intriguing anomalies that are still not completely understood.
Both in crystals and in disordered solids in $d$ dimensions the (vibrational) density of states (DOS) $D(\omega)$ in the low frequency regime --- i.e., on large scales --- is given by the Debye law, $D(\omega)\sim\omega^{d-1}$. However, disordered systems present a nontrivial deviation from Debye's theory at higher frequencies. This motivated a large amount of literature on the general properties of $D(\omega)$ and of the structure factor in the disordered case, from both the numerical and experimental point of view \cite{Phillips1981,*Malinovsky1986,OHern2003}. 

For example, the so-called ``Boson peak'' --- an excess of modes with respect to Debye's prediction --- is a common feature of the DOS of disordered solids. It has been interpreted as a precursor of instability in harmonic regular lattices with spatially fluctuating elasticity \cite{Schirmacher1998,*Schirmacher2006,*Marruzzo2013,*Schirmacher2015} and it seems to be linked to the Ioffe-Regel crossover frequency \cite{Xu2009,Beltukov2013}. It has also been suggested that the Boson peak is simply a smeared version of the van Hove singularity, a well known feature of crystals \cite{Chumakov2011,*Chumakov2014}. A different point of view on this topic came from the study of the dynamic structure factor in supercooled liquids, which has been successfully tackled using Euclidean random matrix theory \cite{Mezard1999,MartinMayor2001,*Ciliberti2003,MartinMayor2000,*Grigera2002,*Grigera2011,Cavagna1999,*Cavagna2000}. From this perspective, the Boson peak phenomenon can be interpreted as a phonon-saddle transition \cite{Parisi2002a,*Grigera2003}. The relation between disorder and the Boson peak is, however, still a matter of debate, alongside other spectral properties of disordered solids.

\begin{figure}
\hfill
\includegraphics[height=0.22\textwidth]{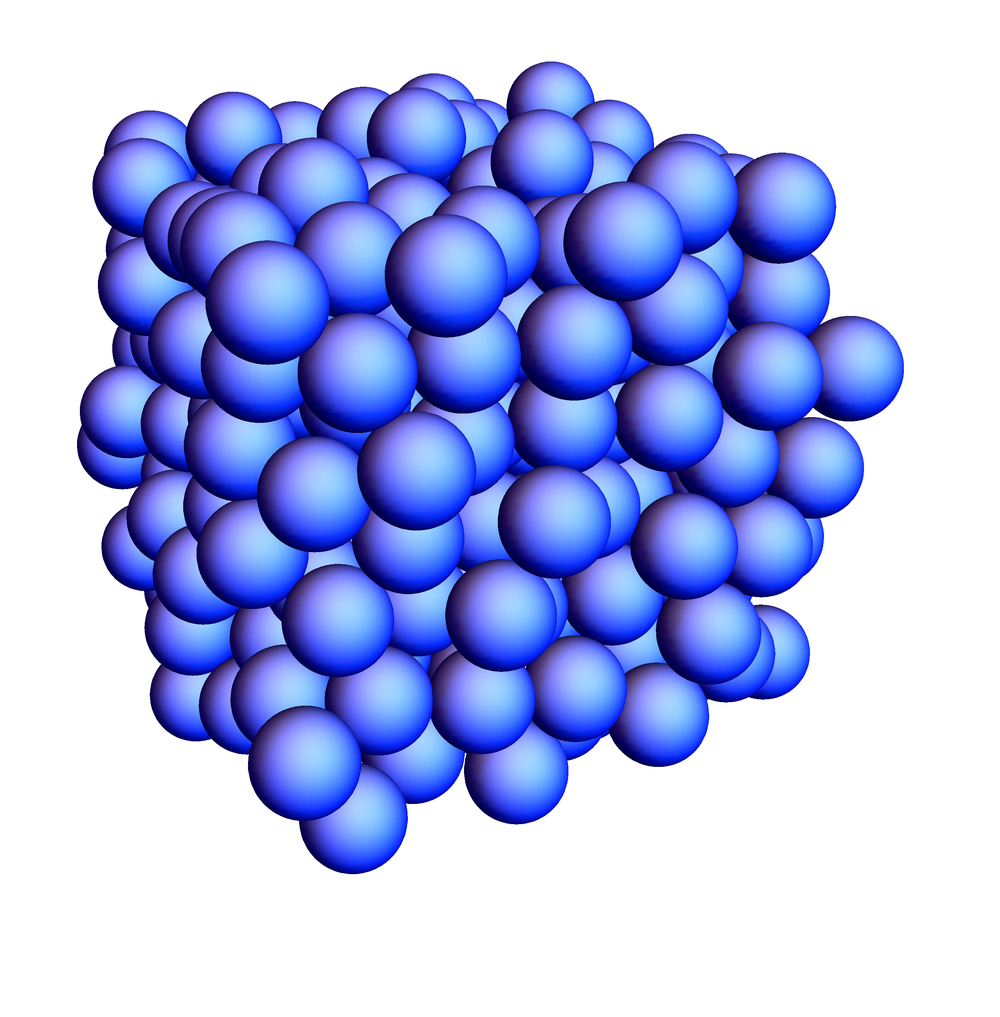}\hfill
\includegraphics[height=0.22\textwidth]{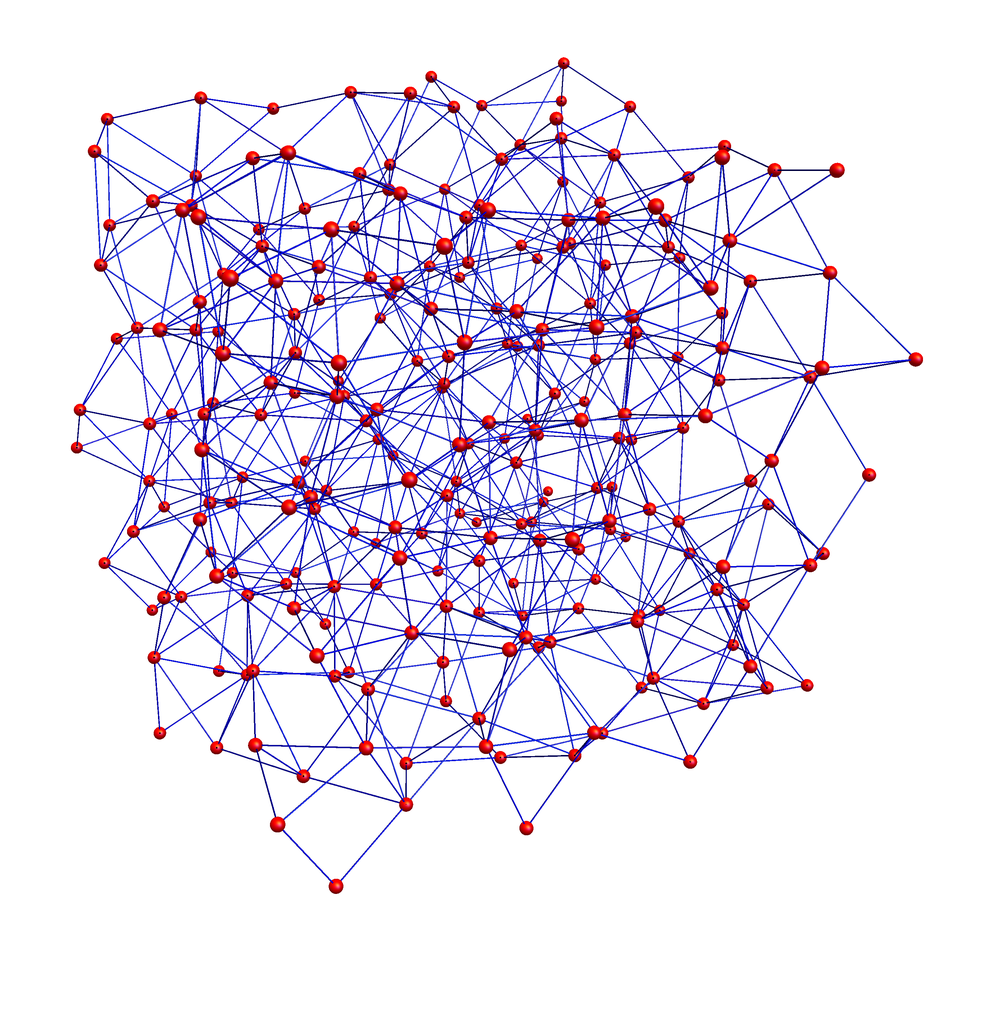}\hfill\,
\caption{Pictorial representation of a small overjammed system ($N=256$) of soft spheres in $d=3$ dimensions with its corresponding contact network. The overjammed configuration has been obtained assuming periodic boundary conditions. To simplify the network figure the contacts across the boundary are not shown.\label{fig:contatti}}
\end{figure}

In the present paper, we want to study the properties of $D(\omega)$ in a mean-field model for soft spheres near the jamming point. The simplest model of a disordered system of soft spheres is an elastic network with some kind of randomness in it. Random elastic networks have a long tradition in the literature. For example, an effective-medium theory (EMT) has been developed for the study of a system of oscillators on a regular lattice of springs with random stiffness  \cite{Feng1985,*Garboczi1985,*Garboczi1985}. In all these models it emerged quite clearly that one of the essential features that strongly affects the properties of the DOS is the average degree of a node in the network, as first observed by Maxwell in his study on the stability of solids \cite{Maxwell1864,Alexander1998}. By means of a constraint counting, Maxwell showed that, given a system of particles in $d$ dimensions, global mechanical stability requires an average number of contacts per particle given at least by $\bar z=2d$, despite the fact that $z=d+1$ contacts on each particle are enough to pin it in a given position.

Applying Maxwell's argument and using a variational approach, a general qualitative picture of $D(\omega)$ in a disordered elastic solid has been obtained in the last decade \cite{Wyart2005,*Wyart2005c,*Xu2007,*Yan2016}. In particular, assuming $\delta z\coloneqq \bar z-2d>0$, it is expected that $D(\omega)$ has a plateau for $\omega\geq \omega_*\propto\delta z$, and that the plateau extends up to the origin for $\delta z\to 0^+$ \cite{OHern2003,Silbert2005}. The frequency $\omega_*$ increases with compression \cite{Wyart2005b}, due to the fact that $\bar z$ increases by consequence as well. The value $\omega_*$ is directly connected to the Boson peak and to an Ioffe-Regel crossover \cite{Xu2009}. Indeed, using EMT, \textcite{DeGiuli2014} found that the Boson peak frequency scales as $\omega_\text{bp}\sim\sqrt{\omega_e\omega_*}$, where $\omega_e$ is a frequency at which strongly-scattered modes appear and which depends on the compressive strain. A numerical study of the contact network of an overjammed system of soft spheres near the jamming point shows that there is a relation between the average number of contacts $\bar z$ and the packing fraction $\varphi$, i.e., $\delta z\propto(\varphi-\varphi_c)^{\sfrac{1}{2}}$ for $\varphi\geq\varphi_c$, $\varphi_c$ being the jamming transition packing fraction \cite{Durian1995,OHern2003,vanHecke2010}. The two variables $\bar z$ and $\varphi$ therefore play an equivalent role. On the other hand, if $\bar z<2d$, the stability condition is violated and the system is hypostatic: an extensive number of zero (floppy) modes appears and $D(\omega)$ has a gap for $0<\omega<\omega_0$ for a certain frequency $\omega_0$ \cite{During2013}. These results suggest quite clearly that, independently from the amount of disorder, stability is determined by two parameters: the average coordination $\bar z$ and the compressive strain applied to the system. Furthermore, on the transition between stability and instability, the frequency of the Boson peak vanishes and its amplitude diverges. 

Despite the fact that the general features of the DOS in the three regimes are well established, the low-frequency properties of $D(\omega)$ for $\delta z>0$ are still a matter of investigation. In this regime, using an EMT approach, \textcite{DeGiuli2014} predicted
\begin{equation}\label{domega}
 D(\omega)\sim\begin{cases}
               \sfrac{\omega^{d-1}}{\omega_*^{\sfrac{d}{2}}},&\omega\ll \omega_e\\
               \sfrac{\omega^2}{\omega_*^2},&\omega_e\ll \omega\ll\omega_*\\
               \text{constant},&\omega\gg\omega_*.
              \end{cases}
\end{equation}
The same behavior has been obtained in the study of the soft perceptron, the simplest possible mean-field model for jamming \cite{Franz2015,*Franz2016}. The lowest frequency behavior in Eq.~\eqref{domega} corresponds to the phonon contribution, which is absent for $d\to+\infty$. What happens in finite dimension if the phonons are removed, however, is a nontrivial question. Indeed, both EMT and the perceptron model, which are mean-field theories, suggest that, for $\omega\to 0$, $D(\omega)\sim\omega^2$ once the Goldstone modes are neglected. On the other hand, \textcite{Gurarie2003} and \textcite{Gurevich2003} predicted a $D(\omega)\sim \omega^4$ scaling as a general behavior of the DOS in random media for $\omega\to 0$ in finite dimension. Numerical evidence is available in favor of both the mean-field \cite{Charbonneau2016} and the finite-dimensional \cite{BaityJesi2015,Lerner2016,Lupo2017,Mizuno2017,Angelani2018} predictions. Different authors dealt with the presence of phonons in finite dimension in different ways, e.g., by a random external field in spin glasses to break translational invariance \cite{BaityJesi2015,Lupo2017}, or, in structural glasses, carefully tuning the system size \cite{Lerner2016}, isolating the localized low-frequency modes \cite{Mizuno2017}, or performing a random pinning \cite{Angelani2018}. How to recover the finite-dimensional-scaling from the infinite-dimensional one is still an open problem. {Moreover, recent studies suggest that the protocol adopted for cooling the system might be relevant in the final low-frequency power-law behavior. In Ref.~\onlinecite{Lerner2017}, for example, it has been shown that $D(\omega)\sim\omega^3$ in glasses obtained quenching from temperatures much higher than the glass transition temperature.}

The low-frequency regime for $\delta z>0$ is interesting also for its localization properties. The presence of (quasi) localized modes in the lower edge of the spectrum, alongside the presence of localized modes in the upper edge \cite{Silbert2009}, has been observed in systems of soft spheres \cite{Laird1991,*Schober1991} but also in the instantaneous normal modes spectrum of low-density liquids \cite{Cavagna1999,*Cavagna2000,Ciliberti2004}. In these cases the localized low-frequency modes tend to hybridize with extended Goldstone modes, becoming weakly localized. 

The presence of localized low-frequency modes is common in many disordered models. For example, localized states appear on the edge of the spectrum in models with disorder on random graphs and Bethe lattices \cite{Biroli1999b,*Biroli2010}.  Localized eigenstates have been found also on the spectrum edges of Euclidean random matrix models on random graphs \cite{Ciliberti2005}. However, this property is out of the reach of mean-field models for jamming having infinite connectivity \cite{Franz2015,*Franz2016}. The presence of localized low-frequency modes is relevant, because they are precursors of instabilities in the unjamming transition and of local rearrangements in sheared glasses \cite{WidmerCooper2008,*WidmerCooper2009,Manning2011}. Once again, the frequency $\omega_*$ plays the role of crossover frequency between the region of extended modes and the region of modes that are localized on few particles, which typically have low coordination \cite{Wyart2010,Xu2010}. Moreover, the delocalization of low-frequency modes increases as $\varphi\to\varphi_c$ and $d$ increases \cite{Charbonneau2016}.

The complexity of the scenario above motivated us to study a mean-field model in which Goldstone modes are absent by construction, that can be treated by the cavity method and that is still reminiscent of the finite dimensionality of real amorphous packings. We consider a tree-like random graph, which is our model for the (equilibrium) contact network in an amorphous packing. The lack of an underlying lattice regularity automatically forbids Goldstone modes. Each vertex in the graph corresponds to a sphere, and each edge is associated to a $d$-dimensional random unit vector joining the centers of two spheres in contact. The Hessian matrix $\bM$ is therefore constructed using this set of random vectors on the graph, and the DOS is computed from the spectrum of $\bM$ and averaging over all realizations. {This model has been investigated by \textcite{Parisi2014} on random regular graphs, and explicit expressions for the first moments of the corresponding DOS on Erd\H{o}s--R\'enyi random graphs are available \cite{Cicuta2018}. 

The model discussed above has been inspired by the one introduced by \textcite{Manning2015}, the so-called ``diagonal-dominant (DD) random matrix model''. In the DD model the Hessian matrix $\bM$ is constructed using an Erd\H{o}s--R\'enyi random graph with average coordination $\bar z$, in such a way that a random scalar quantity is associated to each edge. This model is therefore intrinsically ``one-dimensional''. Isostaticity corresponds to $\bar z=2$ and thus there is no hypostatic regime.} A similar ``one-dimensional'' model has been very recently considered in Ref.~\onlinecite{Stanifer2018}, where the DOS of a ring of springs with random cross bonds has been studied in the presence of disorder in the elastic constants.

Starting from the results of Ref.~\onlinecite{Parisi2014}, in this paper we consider a more general class of graphs that are also locally tree like, a fact that allows us to apply the cavity method to obtain information about both the DOS and the localization properties of the model. The predictions of the cavity method will be compared with the results obtained through an exact diagonalization procedure and the method of moments. 

The paper is organized as follows. In Section~\ref{sec:model} we describe in detail the model under investigation and the methods that we have adopted to solve it. In Section~\ref{sec:risultati} we present our results for three possible cases (hypostatic, isostatic, and hyperstatic regimes). We compare the results obtained with fixed and with fluctuating coordination, stressing the main differences between the two cases. Finally, in Section~\ref{sec:conclusioni} we give our conclusions.

\section{Model and methods}\label{sec:model}
Let us consider a system of $N$ soft spheres in $d$ dimensions, whose centers are in positions $\{\mathbf r_i\}_{i=1,\dots,N}$, $\mathbf r_i=(r_i^\mu)_{\mu=1,\dots,d}$ being a $d$-dimensional vector in the Euclidean space. We assume that the spheres interact by a finite-range repulsive potential $U(x)$ depending on the modulus of their Euclidean distance only. We also assume that there are $N_c$ total ``contacts'' among the spheres, two spheres being in contact if there is a nonzero interaction between them. A given configuration of the spheres can therefore be naturally associated to a contact network, i.e., a graph $\mathcal G=(\mathcal V,\mathcal E)$ with vertex set $\mathcal V$ of cardinality $N$, and edge set $\mathcal E$ of cardinality $N_c$, in such a way that the $i$th sphere corresponds to the vertex $i\in\mathcal V$ and the edge $e=(i,j)$ is an element of $\mathcal E$ if, and only if, the $i$th sphere and the $j$th sphere are in contact (see Fig.~\ref{fig:contatti}). The average coordination number of the graph, i.e., the average number of contacts of each sphere, is given by
\begin{equation}\label{zmedio}
\bar z\coloneqq\frac{2N_c}{N}.
\end{equation} 

Denoting by $\mathbf x_{ij}\coloneqq \mathbf r_i-\mathbf r_j$ the distance between the $i$th sphere and the $j$th sphere, the Hamiltonian of the system depends on the set of distances $\{\mathbf x_{ij}\}_{(i,j)\in\mathcal E}$ only, and it can be written as
\begin{equation}\label{Hamhat}
 \hat{\mathcal{H}}=\sum_{\mathclap{(i,j)\in\mathcal E}}U\left(\|\mathbf r_{i}-\mathbf r_j\|\right)\equiv\sum_{\mathclap{(i,j)\in\mathcal E}}U\left(\|\mathbf x_{ij}\|\right).
\end{equation}
Given a set of \textit{equilibrium positions} of the spheres, we can easily write down a quadratic Hamiltonian function that describes the fluctuations of the system around the given minimum (an inherent structure) by means of a harmonic approximation of the Hamiltonian in Eq.~\eqref{Hamhat}.  Let $\bdelta_i$ be the fluctuation of the $i$th sphere around its equilibrium position $\mathbf r_i$. We assume that, at equilibrium, $\|\mathbf r_{i}-\mathbf r_j\|=\|\mathbf x_{ij}\|=1$ for all $(i,j)\in\mathcal E$ and, moreover, we will neglect the so-called ``initial stress'' contribution \cite{Alexander1998}, that indeed vanishes at jamming. Up to an additive constant and a global multiplicative factor, a quadratic approximation of Eq.~\eqref{Hamhat} gives us
\begin{subequations}\label{Hamiltonian}
\begin{equation}\label{Hamiltoniana}
\mathcal H[\bdelta]\coloneqq 
\sum_{ij}\sum_{\mu,\nu=1}^d \delta_i^\mu M_{ij}^{\mu\nu}\delta_j^{\nu}.
\end{equation}
The element $\mathbf M_{ij}$ of the Hessian matrix $\bM=(\mathbf M_{ij})_{ij}$ is a $d\times d$ matrix given by
\begin{equation}\label{proprieta}
\mathbf M_{ij}=
\begin{cases}
-|\mathbf x_{ij}\rangle\langle\mathbf x_{ij}|&\text{if $(i,j)\in\mathcal E$},\\
\sum\limits_{k\in\partial i}|\mathbf x_{ik}\rangle\langle\mathbf x_{ik}|=-\sum\limits_{k\in\partial i}\mathbf M_{ik}&\text{if $i=j$},\\
\mathbf 0&\text{otherwise}.
\end{cases}
\end{equation}
In the expression above, $\partial i$ is the set of neighbors of the vertex $i$ in the graph, i.e., the set of all the spheres in contact with the sphere $i$. For the sake of brevity, here and in the following we use a bra-ket notation, representing, for example, by $|\mathbf x\rangle$ the vector $\mathbf x\in\mathds R^d$ and by $|\mathbf x\rangle\langle\mathbf y|=(x^\mu y^\nu)_{\mu\nu}$ the outer product.  Observe that the translational invariance constraint
\begin{equation}\label{proprieta2}
 \sum_{k=1}^N M_{ik}^{\mu\nu}=0\qquad\forall i\in\mathcal V,\ \forall \mu,\nu=1,\dots,d
\end{equation}\end{subequations}
is satisfied. The DOS $D(\bM;\omega)$ of the system can be obtained directly from the spectral density $\varrho(\bM;\lambda)$ of the Hessian matrix in Eqs.~\eqref{Hamiltonian}, by means of the change of variable $D(\bM;\omega)=2\omega\varrho(\bM;\omega^2)$. In particular, the vibrational DOS $D(\bM;\omega)$ is a comb of $Nd$ Dirac deltas,
\begin{equation}
 D(\bM;\omega)=\frac{1}{Nd}\sum_{k=1}^{Nd}\delta(\omega-\omega_k)\equiv 2\omega\varrho(\bM;\omega^2),
\end{equation}
where $\omega_k=\sqrt{\lambda_k}$, $\lambda_k$ being the $k$th eigenvalue of the dynamical matrix $\bM$. Observe that in the system described by the Hamiltonian in Eqs.~\eqref{Hamiltonian}, $d$ zero modes are always present, due to the fact that the translational invariance allows $\bdelta_i\mapsto \bdelta_i+\boldsymbol\lambda$ for any $\boldsymbol\lambda\in\mathds R^d$, and therefore there will always be a $\delta(\omega)/N$ contribution in the DOS.

To introduce and study the effects of randomness, we adopt a mean-field approximation \cite{Parisi2014}. We first suppose that the $N_c$ quantities $\mathbf x_{ij}$ appearing in Eqs.~\eqref{Hamiltonian} are independently generated random $d$-dimensional Gaussian unit vectors. Moreover, we suppose that the graph $\mathcal G$ is a random graph in which the coordination $z$ is distributed according to certain probability distribution $p_k$ such that $\Pr(z=k)=p_k$ for $k\in\mathds N$. For each value of $d$, we require that the coordination number $z_i$ of the $i$th vertex always satisfies the local stability condition $z_i\geq d+1$, and therefore $p_k=0$ for $k<d+1$. The translational invariance constraint in Eq.~\eqref{proprieta2} appears to be crucial in a random matrix model for the vibrational DOS of a disordered solid \cite{Manning2015} and it will be preserved. In this way randomness is introduced both in the edge weights and in the topology of the graph. We are interested in the properties of the DOS in the thermodynamical limit $N\to+\infty$ and keeping $\bar z$ constant.

In this paper, we will study two different random graph ensembles, always assuming $d=3$, if not otherwise specified.

We will first consider random regular graphs, i.e., graphs having $p_k=\delta_{k,\bar z}$. We will denote this model by $\mG_{\bar z,0}$. Following Ref.~\onlinecite{Parisi2014}, we have analyzed the three cases $\bar z=5$, $\bar z=6$, and $\bar z=7$, corresponding to a hypostatic, isostatic, and hyperstatic system, respectively.

We have then considered a second, more realistic class of graphs, in which fluctuations in the coordination are allowed. In an element of this second class of graphs, the coordination of each vertex $i$ is given by $z_i=z_0+\zeta_i$, where $z_0\geq d+1=4$ is a constant and $\zeta_i$ is a Poisson random variable having mean $\bar\zeta$. It follows that, in this case, $p_k=\bar\zeta^{k-z_0}\frac{\e^{-\bar\zeta}}{(k-z_0)!}$ for $k\geq z_0$, and zero otherwise. An element of this class can be thought of as an Erd\H{o}s--R\'enyi random graph ``superimposed'' on a random regular graph. We will denote this model by $\mG_{z_0,\bar{\zeta}}$. In our analysis, we have chosen $z_0$ and $\bar \zeta$ in such a way that either $\bar{z}<6$, or $\bar{z}=6$, or $\bar{z}>6$, corresponding to the hypostatic, isostatic, and hyperstatic case, respectively. This model reproduces in a reasonable way the real coordination distribution of sphere packings near jamming \cite{Charbonneau2015,Jin2018} and allows us to consider the effects of fluctuating coordination\footnote{{Observe here that, once the value $\bar z$ for the ensemble is fixed, for finite $N$, each graph has an average coordination $z_{\text{av}}=\sfrac{1}{N}\,\sum_{i=1}^N z_i$ that fluctuates around $\bar z$, due to finite-size effects. This is particularly important for $\bar z=6$, because in this case, if no constraint is imposed on the graph, at finite $N$ an instance will be in general either hypostatic or hyperstatic. To avoid this problem, we have accepted only graphs having $|z_\text{av}-\bar z|\leq \sfrac{1}{M}$ with $M\gg N$ for each analyzed $N$, in such a way that fluctuations were so small that the number of zero modes was exactly equal to the expected one.}}. 

Both types of random graphs are locally tree-like and the models combine a mean-field approximation (the random graph topology) with the finite number of degrees of freedom of each contact, which is reminiscent of a finite dimensionality.  

\subsection{Density of states and the cavity method approach}
As usual in the study of disordered systems, we are interested in the properties of our model averaged over disorder, and in particular in the average DOS, namely, 
\begin{equation}\label{gomega}
 D(\omega)\coloneqq\EE{D(\bM;\omega)}=2\omega \EE{\varrho(\bM;\omega^2)}\eqqcolon 2\omega\varrho(\omega^2),
\end{equation}
where the average $\EE{\bullet}$ is performed over all instances of the problem. After some manipulations of the Dirac deltas, it can be shown \cite{MartinMayor2001,*Ciliberti2003} that the DOS $D(\omega)$ can be written as
\begin{equation}
 D(\omega)= -\lim_{\varepsilon\to 0}\lim_{N\to \infty}\frac{2\omega}{Nd\pi}\EE{\tr\Imm\boldsymbol{\mathsf R}(\omega^2+i\varepsilon) }
\end{equation}
where we have introduced the resolvent
\begin{equation}
 \boldsymbol{\mathsf R}(\lambda)\coloneqq\frac{1}{\lambda \boldsymbol{\mathsf I}_{Nd}-\bM}.
\end{equation}
Here and in the following $\boldsymbol{\mathsf I}_{k}$ is the $k\times k$ identity matrix. In this approach we make, as usual, the assumption that the quantity $D(\omega)$ is self-averaging. We denote by $\mathbf R_{ij}$ the $d\times d$ submatrix of $\boldsymbol{\mathsf R}$ corresponding to the couple of (not necessarily distinct) sites $(i,j)$. Assuming that no vertex plays a special role in the ensemble of realizations, the DOS can be expressed in terms of the averaged trace of the local resolvent $\mathbf R_{ii}$, i.e.,
\begin{equation}
 D(\omega)
 =-\lim_{\varepsilon\to 0}\lim_{N\to \infty}\frac{2\omega}{d\pi}\EE{\tr\Imm\mathbf{R}_{ii}(\omega^2+i\varepsilon)}.
\end{equation}

Before proceeding further, let us comment on some properties of the Hessian matrix under analysis. For each realization of our system, the matrix $\bM$ has dimension $dN\times dN$, but it has rank $N_c=\sum_{i\in\mathcal V} z_i/2=N\bar z/2$, where $z_i$ is the coordination number of the $i$th vertex. Therefore, if $\bar z<2d$, there are $N\left(d-\sfrac{\bar z}{2}\right)$ zero modes. In that case, a contribution $\left(1-\sfrac{\bar z}{2d}\right)\delta(\omega)$ to the DOS $D(\omega)$ appears, corresponding to a singularity in the trace of the local resolvent for $\lambda\to 0$ of the type
\begin{equation}\label{trhatR}
 \EE{\tr\mathbf R(\lambda)}=\frac{2d-\bar z}{2\lambda}+o\left(\frac{1}{\lambda}\right).
\end{equation}

By the same argument, moreover, no singularity is expected for $\lambda\to 0$ for $\bar z=2d$. This is nothing other than Maxwell's criterion, which implies instability for $\bar z<2d$ due to the presence of an extensive number of zero modes.

A typical approach for the solution of Eq.~\eqref{gomega} in the thermodynamical limit is the cavity method \cite{mezard1987spin,AbouChacra1973,Cizeau1994}, which is exact on a Bethe lattice and can be applied when the underlying topology is a tree like graph. Using this approach, it can be proved (see Appendix~\ref{app:ricorsiva}) that the local resolvent $\mathbf R$ satisfies \textit{in probability} the equation
\begin{subequations}\label{ricorsive}
 \begin{equation}\label{ricorsiva2}
 \mathbf R(\lambda)\stackrel{\text{prob}}{=}\left[\lambda\bI_d+\sum_{k=1}^{z}\frac{|\mathbf x_k\rangle\langle\mathbf x_k|}{1+\langle\mathbf x_k|\mathbf G_k(\lambda)|\mathbf x_k\rangle}\right]^{-1},
\end{equation}
where $z$ is distributed according to $p_k$, the degree distribution of the graph, and $\{\mathbf x_k\}_{k=1,\dots,z}$ are $z$ random Gaussian unit vectors in $d$ dimensions. The $\{\mathbf G_k\}_{k=1,\dots,z}$ are $z$ local \textit{cavity fields} satisfying a similar equation
\begin{equation}\label{ricorsiva}
\mathbf G(\lambda)\stackrel{\text{prob}}{=}\left[\lambda\bI_d+\sum_{k=1}^{\eta-1}\frac{|\mathbf x_k\rangle\langle\mathbf x_k|}{1+\langle\mathbf x_k|\mathbf G_k(\lambda)|\mathbf x_k\rangle}\right]^{-1},
\end{equation}
the main difference being the fact that the random variable $z$ is replaced by the random variable $\eta$, which is distributed with probability distribution \cite{mezard2009information} 
\begin{equation}
\hat p_\eta=\frac{\eta p_\eta}{\sum_k kp_k}.
\end{equation}
\end{subequations}
Eqs.~\eqref{ricorsive} provide a recipe for the numerical evaluation of $\EE{\tr\Imm\mathbf R}$ through a population dynamics algorithm \cite{Mezard2001}. We therefore tackled the problem of the DOS of our model both numerically solving Eqs.~\eqref{ricorsive}, through exact diagonalization (ED) via the implicitly restarted Lanczos method~\cite{arpack} and through the method of moments~\cite{Cyrot-Lackmann1967,*Gaspard1973,*Lambin1982,*Jurczek1985, *Benoit1992,*Villani1995} (see Appendix~\ref{app:mom}). In particular, using ED we obtained the lowest part of the average spectrum, calculating the 50 lowest eigenmodes (or 100 for the smaller system sizes), whereas the rest of it has been obtained from the matrix $\bM$ using the method of moments. 

\begin{figure}
\includegraphics[width=0.7\columnwidth]{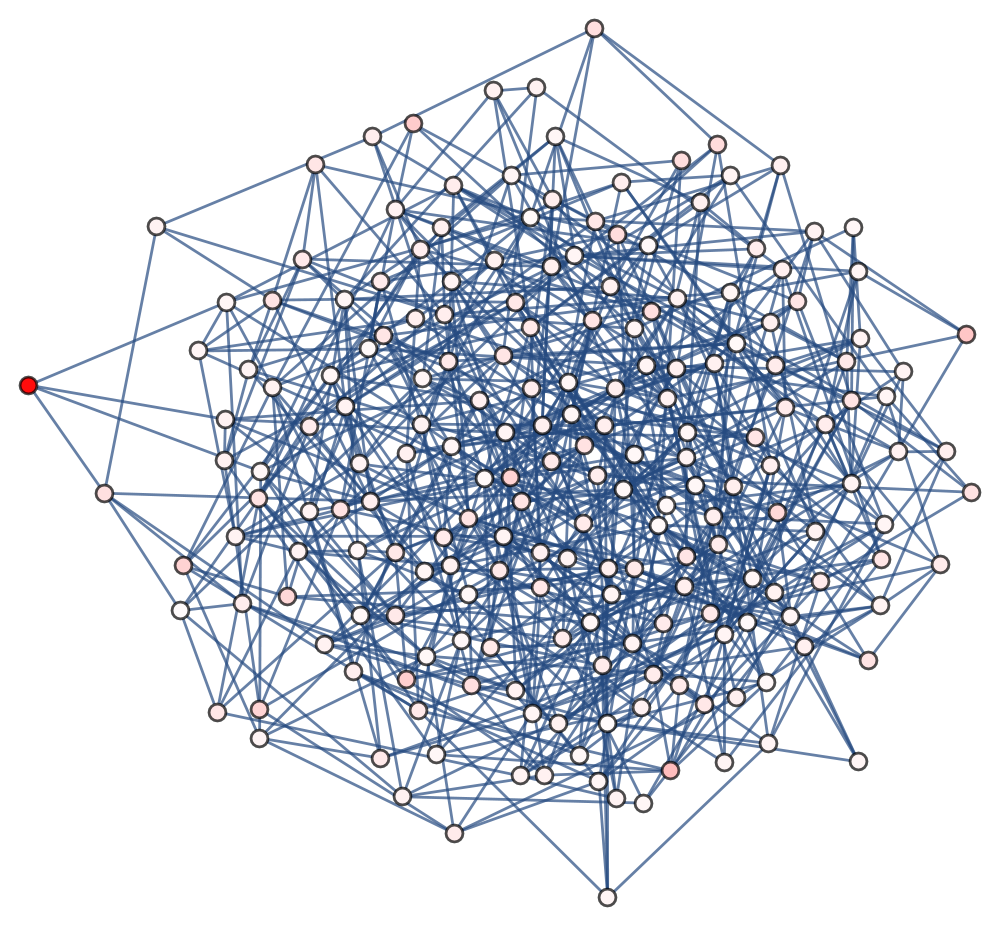}
\caption{An instance of the $\mG_{4,3}$ model for $N=200$ with a low-frequency eigenmode represented on it. The intensity of the color is proportional to the amplitude of the corresponding eigenmode on each site. It is evident that the eigenmode is localized on a site with $z=4$, the lowest possible coordination. \label{fig:soffice}}
\end{figure}

\subsection{Eigenvectors localization}

We also investigate the localization phenomenon near the band edges in the model proposed above. Let us denote by $|{\boldsymbol{\mathsf k}}\rangle$ the eigenmode of the matrix $\bM$ corresponding to the eigenvalue $\lambda_k$, $\bM|{\boldsymbol{\mathsf k}}\rangle=\lambda_k|{\boldsymbol{\mathsf k}}\rangle$, and by $|{\mathbf k_i}\rangle$ its projection on the site $i$. In this paper, we will always assume that the eigenvectors are labeled in such a way that $k<k'\Rightarrow \lambda_k\leq \lambda_{k'}$. We use as an indicator for the localization of the eigenvector $|{\boldsymbol{\mathsf k}}\rangle$ the inverse participation ratio (IPR)
\begin{equation}
 Y_k\coloneqq\frac{\sum_{i=1}^N\left|\langle \mathbf k_i|\mathbf k_i\rangle\right|^2}{\left(\sum_{i=1}^N\langle \mathbf k_i|\mathbf k_i\rangle\right)^2}.
\end{equation}
The IPR scales as $O(1)$ if the eigenvector $|{\boldsymbol{\mathsf k}}\rangle$ is localized, or $O\left(N\right)$ if it is delocalized. The quantity above can be evaluated once the eigenvectors are known from an ED procedure on a given instance of the problem. 
To average over disorder we calculate the quantity 
\begin{equation}\label{eq:avIPR}
Y(\EE{\omega_k})\coloneqq\EE{Y_k},
\end{equation}
i.e., the average of the participation ratio of the $k$th eigenmode as a function of the corresponding average frequency.

We can also study the localization properties of the eigenvectors with the cavity method, introducing, among the many possibilities \cite{Parisi2014}, the quantity
\begin{equation}\label{iprcm}
\begin{split}
\hat Y(\omega)&\coloneqq \frac{\EE{\sum \limits_{k\colon \lambda_k\sim\omega^2}\left|\langle \mathbf k|\mathbf k\rangle\right|^2}}{\left\{\EE{\sum \limits_{k\colon \lambda_k\sim\omega^2}\langle \mathbf k|\mathbf k\rangle}\right\}^2}\\
&=\lim_{\varepsilon\to 0}\left.\frac{\EE{\tr\left(\mathbf R^\dag(z)\mathbf R(z)\right)^2}}{\left\{\EE{\tr\Imm\mathbf R(z)}\right\}^2}\right|_{z=\omega^2+i\varepsilon}.
\end{split}
\end{equation}
The last equality allows us to estimate $\hat Y(\omega)$ using the cavity method. It shares the same properties of $Y(\omega)$, i.e., diverges in the localized region and it is $O(1)$ in the delocalized region.

The localization and delocalization properties can also be detected using a different approach. The expected value of the square of
\begin{equation}
\eta_\omega\coloneqq\left.\Imm\tr\mathbf R(\omega^2+i\varepsilon)\right|_{\varepsilon\to 0} 
\end{equation}
should diverge in the localized regime and therefore it can be seen as a localization indicator as well. The divergence of $\EE{\eta_\omega^2}$ can be evaluated either directly or, as we will see below, studying the probability density of $\eta_\omega$ \cite{Parisi2014}. 

\section{Results}\label{sec:risultati}
In this Section we present our results for the DOS $D(\omega)$ and the IPR near the lower band edge. The tools that we use are the ones described in Section~\ref{sec:model}. We also consider the cumulative function
\begin{equation}
 \Phi(\omega)\coloneqq\int_0^\omega D(u)\dd u.
\end{equation}
We will distinguish between the hypostatic, isostatic and hyperstatic cases. As previously stated, in all cases under consideration, we have assumed $d=3$.

\subsection{The hypostatic case} 
\begin{figure*}
  \subfloat[\label{fig:DOSfixed5all}]{\includegraphics[height=0.59\columnwidth]{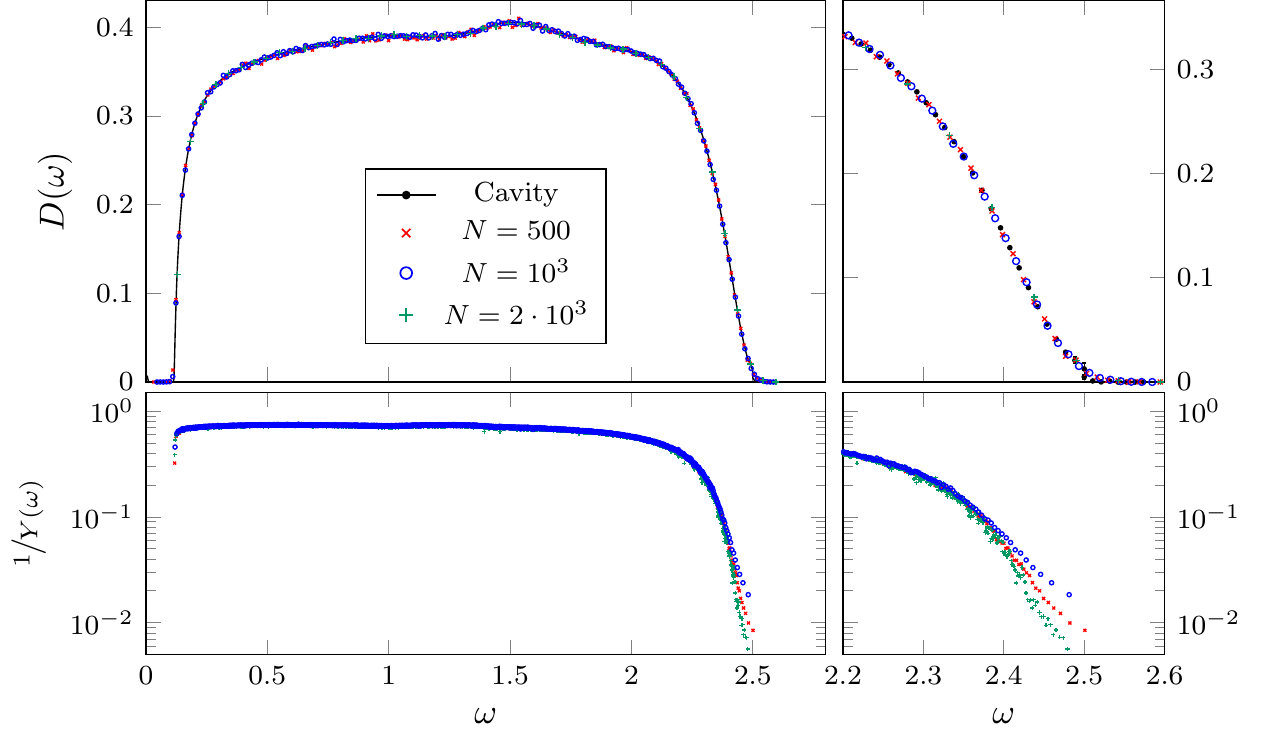}}\hfill
  \subfloat[\label{fig:DOS5all}]{\includegraphics[height=0.59\columnwidth]{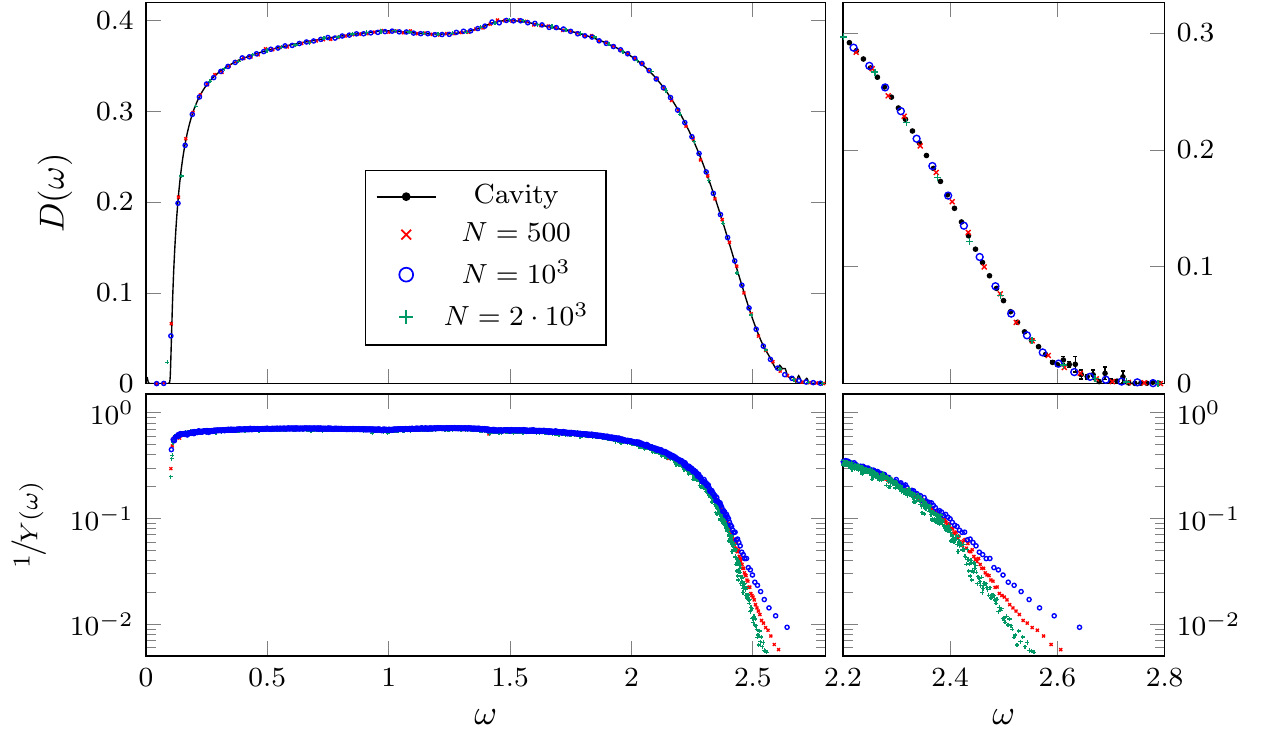}}\hfill
 \caption{DOS and participation ratio [Eq. \eqref{eq:avIPR}] for the $\mG_{5,0}$ model (a) and the $\mG_{4,1}$ model (b) in the hypostatic case, $\bar z=5$, using the cavity method (black dots) and ED (color symbols). The numerical integration of the cavity method equations has been performed using $\varepsilon=10^{-8}$ and a population of at least $10^7$ fields for $\omega<0.2$, and a population of $10^6$ fields for the rest of the interval. The ED results were obtained for $N=500$ (red squares), $N=1000$ (blue circles), and $N=2000$ (green crosses).}
\end{figure*}

\begin{figure}
\includegraphics[width=\columnwidth]{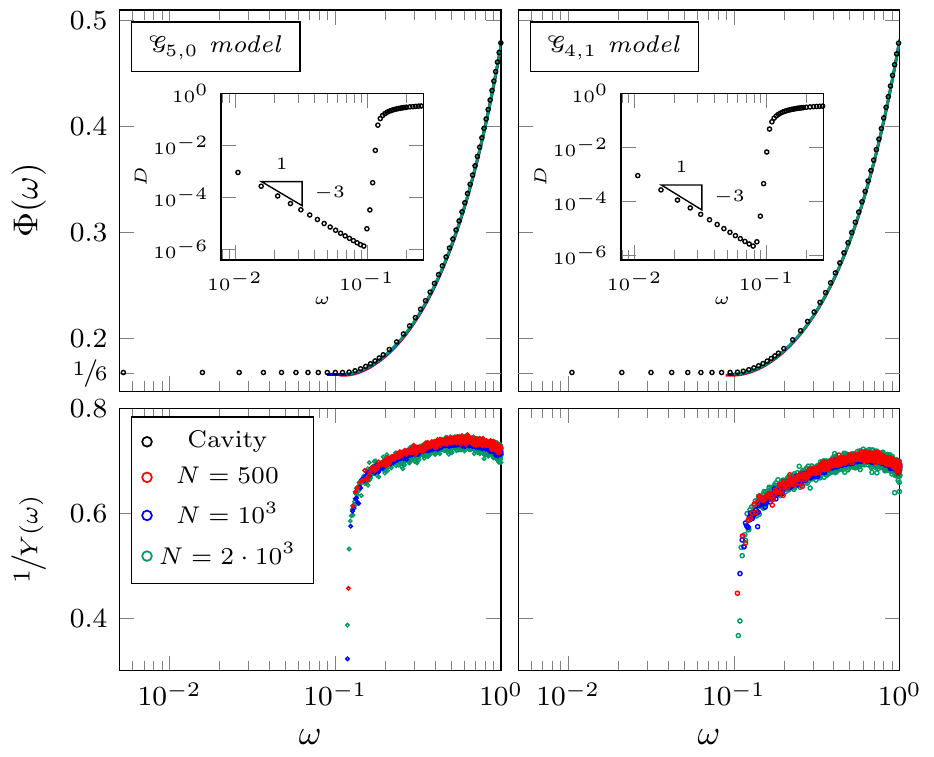}
\caption{Detail of the DOS $D(\omega)$, the cumulative function $\Phi(\omega)$ and the participation ratio $Y(\omega)$ [Eq. \eqref{eq:avIPR}] at low frequencies for the $\mG_{5,0}$ model (left) and the $\mG_{4,1}$ model (right) using the cavity method (black) and ED (color).\label{fig:DOS5head}}
\end{figure}

\paragraph{Density of states.} Let us start from the $\mG_{5,0}$ model, and therefore in the hypostatic regime. A hypostatic network is a good model for the so-called ``floppy materials'', such as dense suspensions, gels, and glasses of low valence elements, which show an abundance of zero modes. In Fig.~\ref{fig:DOSfixed5all} we compare the results of the cavity method and ED on the full spectrum for small sizes of the system for $\bar z=5$, finding an excellent agreement. Both the ED and the cavity results suggest that a gap is present for $\omega<\omega_0\approx 10^{-1}$, as expected in floppy materials \cite{During2013}. The detail of the small frequency regime is shown in Fig.~\ref{fig:DOS5head}. Note that, for $\omega<\omega_0$ and finite $\varepsilon$, a small, nonzero contribution is predicted by the cavity method (see the inset). This contribution is however related to the unavoidable finiteness of the value of $\varepsilon$ adopted in the numerical calculation to solve Eqs.~\eqref{ricorsive}. More specifically, the presence of zero modes implies that a Dirac delta appears in the origin in $D(\omega)$. The finiteness of $\varepsilon$ causes a smoothening of the Dirac function that, in absence of any other contribution --- i.e., in the gap region for $\varepsilon\ll\lambda\ll\omega_0^2$ --- gives a $\overline{\tr\Imm\mathbf R(\lambda)}\sim \varepsilon \lambda^{-2}$ scaling near the origin. The ED and cavity method results have been superimposed. As expected, a zero density is found for $\omega<\omega_0$, whereas the two methods are in agreement for $\omega>\omega_0$.

\begin{figure}
\includegraphics[width=0.9\columnwidth]{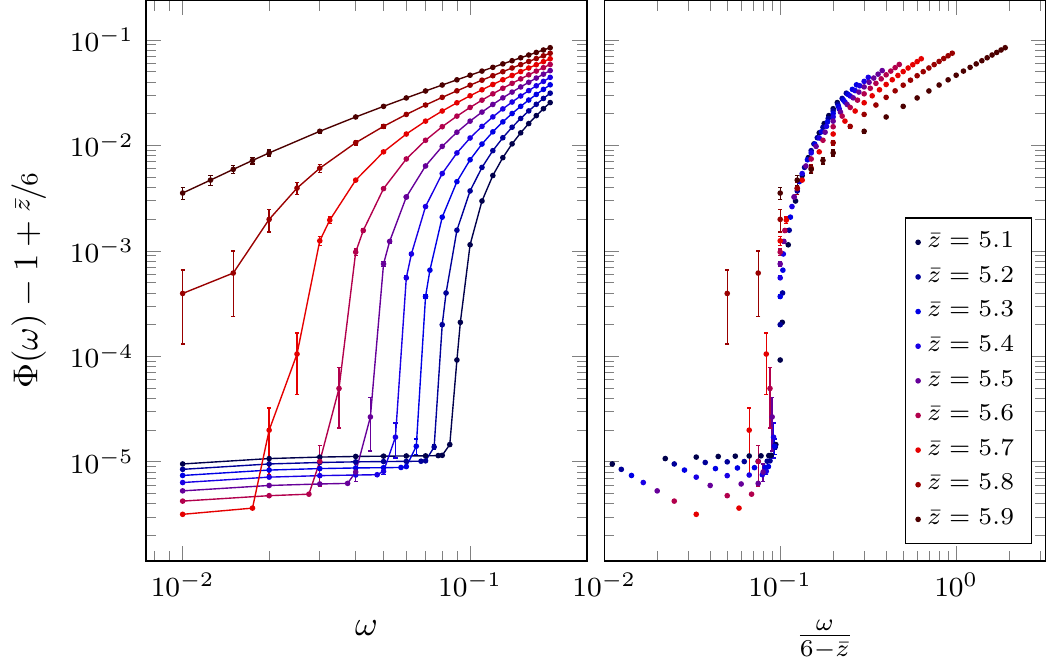}
 \caption{Cumulative function $\Phi$ obtained using the cavity method in the $\mG_{4,\bar\zeta}$ model for different values of $\bar z=4+\bar \zeta<6$. We observe that the value of $\omega_0$ decreases and the gap closes as soon as $\bar z\to 6$. The smooth lines are represented as guides for the eye. On the right panel, the same data are plotted with the $x$ axis rescaled by $6-\bar z$ (the distance from the isostatic transition). \label{fig:5to6}}
\end{figure}

A qualitatively similar result has been obtained for the $\mG_{4,1}$ model, where we find again a gap for $\omega<\omega_0\approx 10^{-1}$ (see Fig.~\ref{fig:DOS5all} and the detail in \ref{fig:DOS5head}). The value of the frequency $\omega_0$ in the $\mG_{4,1}$ model appears to be very close to the one found in the $\mG_{5,0}$ model, showing a weak dependence on the details of the model other than the value $\bar z$. As anticipated, we expect that $\omega_0\to 0$ as $\bar z\to 2d=6$. Taking advantage of the fact that in the $\mG_{4,\bar\zeta}$ model we can smoothly vary $\bar z$, we have computed by cavity method the DOS for $5<\bar z<6$ [see Fig.~\ref{fig:5to6} (left)] and indeed we have observed that $\omega_0$ decreases as $\bar z$ increases, and the gap closes for $\bar z\to 6$. Moreover, the scaling $\omega_0\propto 2d-\bar z$, predicted by \textcite{During2013}, holds in our model [see again Fig.~\ref{fig:5to6} (right)].

\paragraph{Localization properties.} 
Both the $\mG_{5,0}$ model and the $\mG_{4,1}$ model present the same localization features. With reference to Figs.~\ref{fig:DOSfixed5all} and~\ref{fig:DOS5all}, the ED results suggest the presence of a localized region in the upper edge of the spectrum. The participation ratio $Y^{-1}(\omega)$ becomes indeed infinitesimal slightly before the DOS goes to zero (a fact that is more evident in the $\mG_{4,1}$ model), and, moreover, it scales as $O(\sfrac{1}{N})$ for $\omega\gtrsim 2.4$. In the low-frequency regime, instead, $Y(\omega)$ remains $O(1)$ for all the considered sizes up to the lower band edge, suggesting that no mobility edge is present and all eigenstates in the lower part of the spectrum are delocalized.

\subsection{The isostatic case} 

\begin{figure*}
  \subfloat[\label{fig:DOSfixed6all}]{\includegraphics[height=0.58\columnwidth]{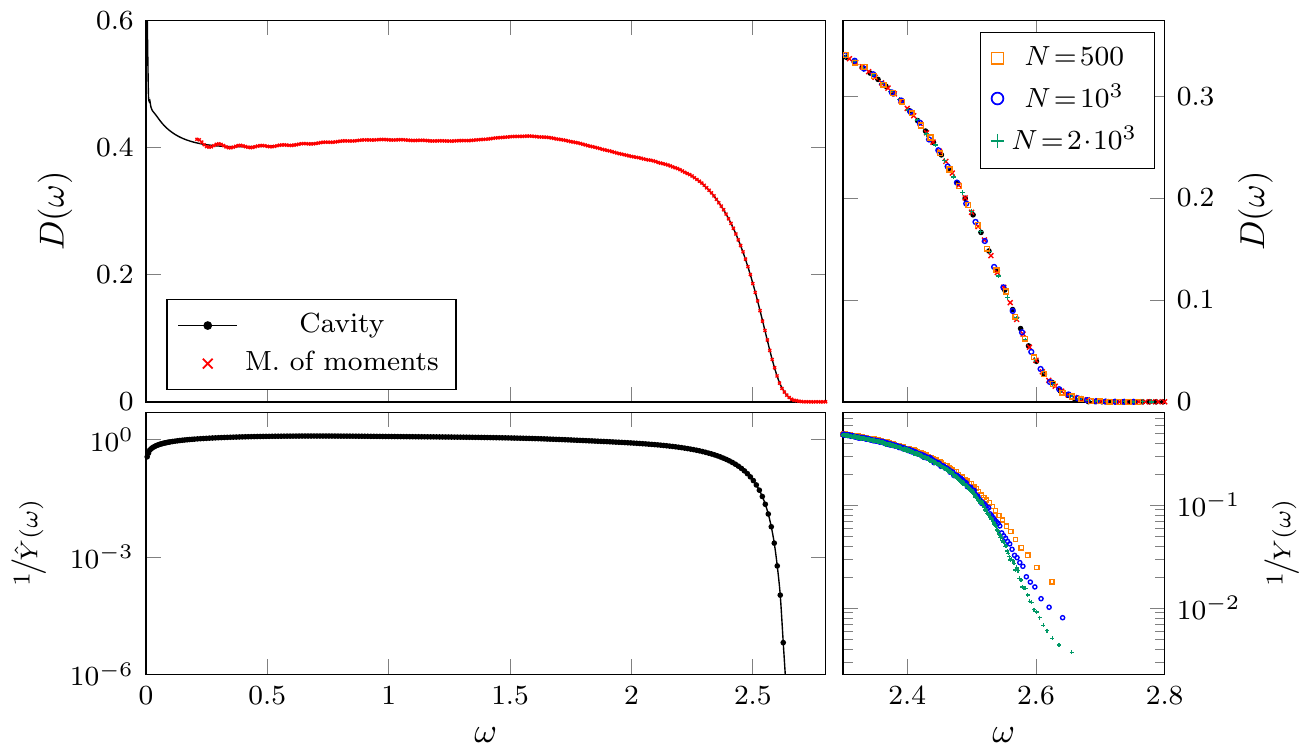}}\hfill
 \subfloat[\label{fig:DOS6all}]{\includegraphics[height=0.58\columnwidth]{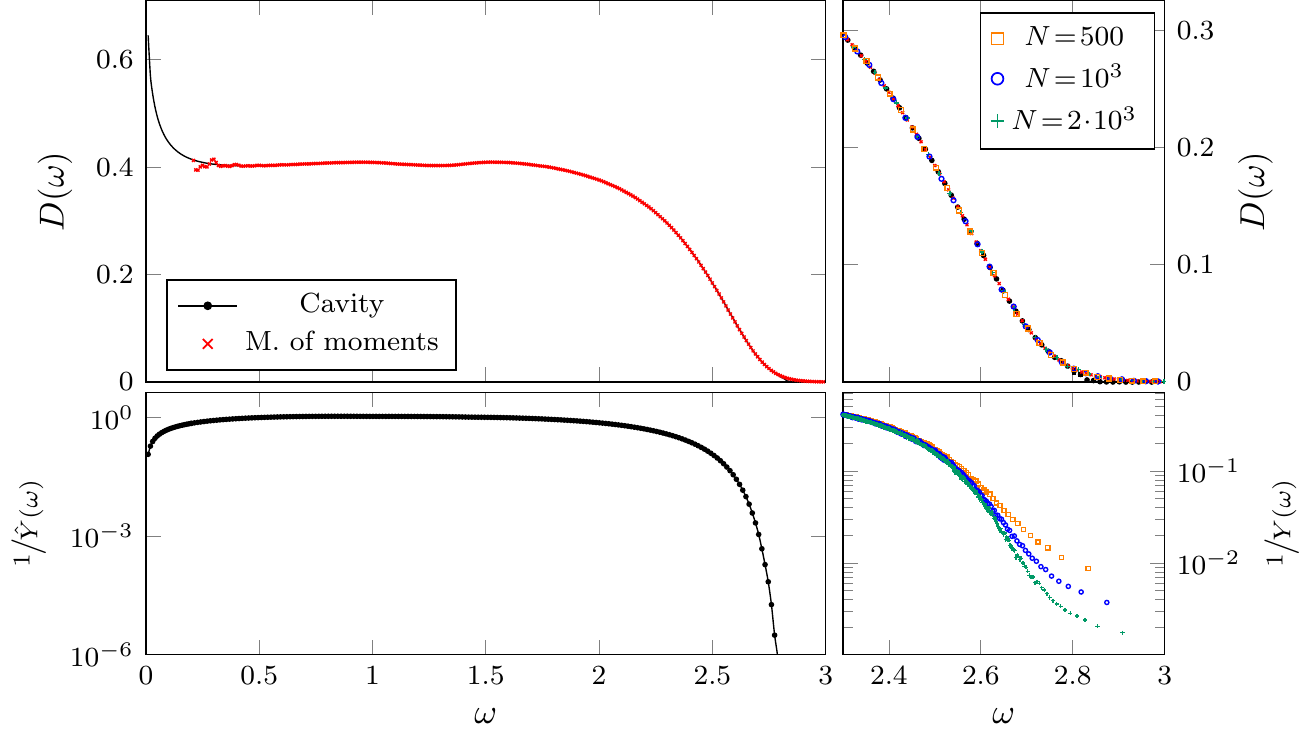}}\hfill
 \caption{DOS, localization indicator $\hat Y$, and participation ratio $\sfrac{1}{Y}$ for the $\mG_{6,0}$ model (a) and the $\mG_{4,2}$ model (b) in the isostatic case, $\bar z=6$, using the cavity method (black), the method of moments (red), and ED. The numerical integration of the cavity method equation has been performed using $\varepsilon=10^{-8}$ and a population of at least $10^7$ fields for $\omega<0.2$, and a population of $10^6$ fields for the rest of the interval. For the method of moments, $50$ moments were used, averaging over $25$ instances of a matrix $\bM$ with $N=10^6$. The method of moments gives very highly fluctuating results for $\omega<0.1$ that have not been represented. ED results for this region are shown in Fig.~\ref{fig:DOS6head}.}
\end{figure*}

\begin{figure}
\includegraphics[width=\columnwidth]{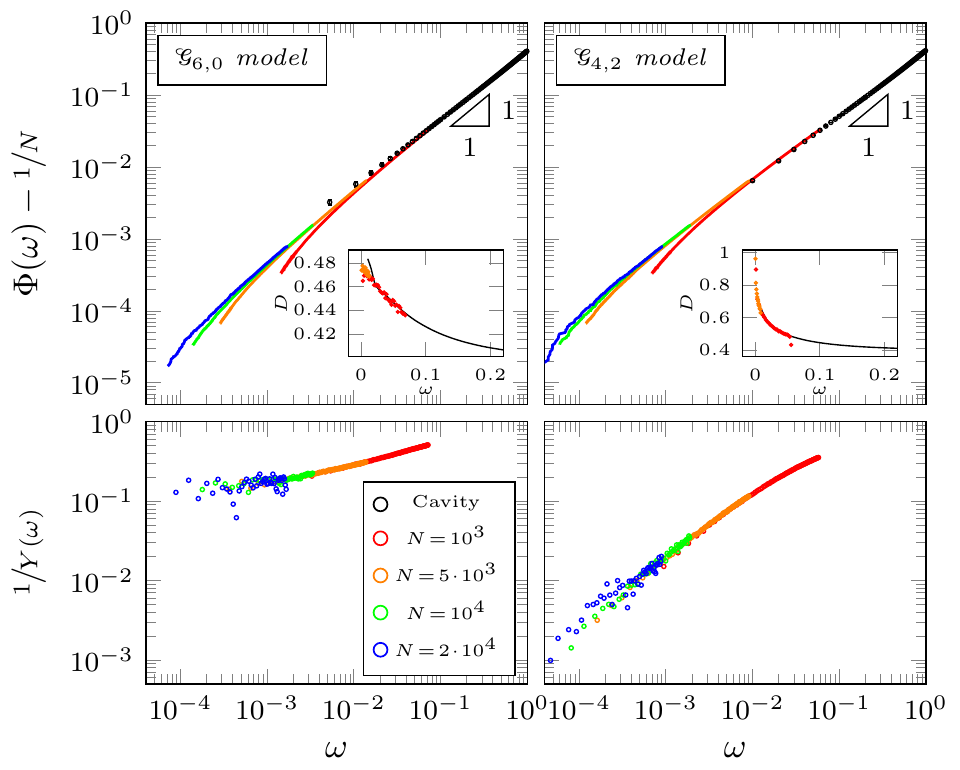}
\caption{Detail of the DOS $D(\omega)$, the cumulative function $\Phi(\omega)$ and the participation ratio at low frequency for the $\mG_{6,0}$ (left) and the $\mG_{4,2}$ (right) models. The results were obtained using ED (color) and the cavity method (black dots)\label{fig:DOS6head}.}
\end{figure}

\paragraph{Density of states.} Let us now consider our model on a random regular graph with $\bar z=6$. As expected from the constraint counting argument, in the $\mG_{6,0}$ model there is no gap and $D(\omega)$ shows a plateau up to low values of $\omega$ (see Figs.~\ref{fig:DOSfixed6all} and~\ref{fig:DOS6head}). A constant $D(\omega)$ for small values of $\omega$ implies that $\varrho(\lambda)\sim \lambda^{-\sfrac{1}{2}}$ for $\lambda \to 0$ and that $\Phi(\omega)\propto\omega$ for $\omega\to 0$. These properties have been verified numerically, and the ED results are compatible with our cavity prediction, as shown in Fig.~\ref{fig:DOS6head}. Both in $D(\omega)$ and in $\hat Y^{-1}(\omega)$ there is, however, an anomalous behavior near $\omega=0$. Both the cavity and the ED results suggest the presence of an integrable singularity in the DOS that is compatible with a logarithmic divergence. Note that it can be proved that no singularity is present in the model for $d\to +\infty$ \cite{Parisi2014}.

Similar results have been obtained in the $\mG_{4,2}$ model, as we show in Fig.~\ref{fig:DOS6all} and in Fig.~\ref{fig:DOS6head}. This suggests that the isostaticity condition $\bar z=2d=6$ is enough to guarantee that there is no gap in the DOS, irrespective of the presence of local fluctuations in the value of $z$. As in the $\mG_{6,0}$ model, for very small values of $\omega$ both methods indicate the presence of an integrable singularity in the origin in the DOS, which in this case appears to be of the type $D(\omega)\sim d_0+\sfrac{d_1}{\omega^\beta}$ for some constants $d_0$ and $d_1$ and with $\beta\approx 0.3$ (see Fig.~\ref{fig:DOS6head}). 

A more precise analysis of this singularity is not possible with the quality of the data that we have in the $\omega>0.01$ range.

\paragraph{Localization properties.} As in the hypostatic case, the participation ratio $\sfrac{1}{Y}$ scales as $O(\sfrac{1}{N})$ for $\omega\gtrsim 2.5$ in the $\mG_{6,0}$ and $\mG_{4,2}$ models (see Figs.~\ref{fig:DOSfixed6all} and~\ref{fig:DOS6all}), suggesting that localized states are present above this threshold. Near the lower band edge we find a value of the IPR that is larger than that of the bulk, yet does not scale with the size of the system. In particular, in the $\mG_{6,0}$ model the IPR increases by a factor $10$ for $\omega\to 0$ for all considered sizes of the system (see Fig.~\ref{fig:DOS6head}), whereas its growth is more evident in the $\mG_{4,2}$ model, where it increases by three orders of magnitude (see Fig.~\ref{fig:DOS6head}) without, however, showing any scaling with $N$. The low-frequency eigenvalues are therefore still delocalized, but the larger IPR is a signal of an avoided localization transition at $\omega=0$.

\begin{figure}
 \includegraphics[width=\columnwidth]{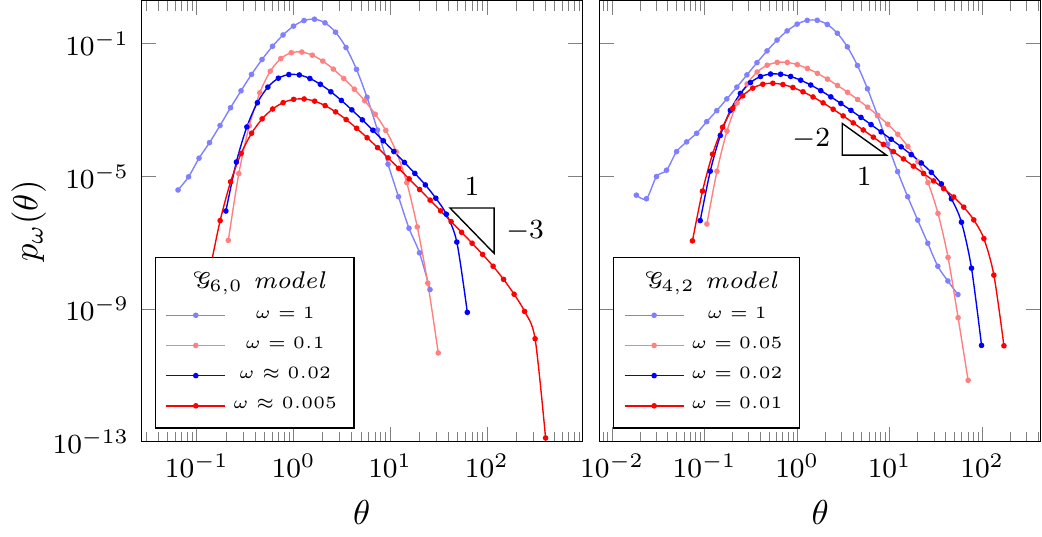}
 \caption{Distribution of the imaginary part of the resolvent $\theta_\omega\coloneqq\omega\eta_\omega$ in the isostatic case for both the $\mG_{6,0}$ model (left) and the $\mG_{4,2}$ model (right).\label{fig:istogrammi6}}
\end{figure}

The incipient localization at very low frequencies can be detected studying the distribution of the imaginary part of the local resolvent $\eta_\omega$ for different values of $\omega$, as described in Section~\ref{sec:model}.  Due to the fact that in this case $\EE{\eta_\omega}\sim\sfrac{1}{\omega}$ for $\omega\to 0$, in Fig.~\ref{fig:istogrammi6} we plot the distribution $p_\omega(\theta)$ for $\theta \coloneqq \omega\eta_\omega$ for different values of $\omega$. In the $\omega\to 0$ limit, a fat tail appears in the $\mG_{6,0}$ model, and in particular we find $p_\omega(\theta)\sim\theta^{-3}$. Such a tail would imply a divergent $\EE{\eta_\omega^2}$ and therefore localization. The exponent can be justified by means of a qualitative argument. In Eq.~\eqref{ricorsiva2} the imaginary part of the resolvent $\eta_\omega$ is related to the inverse of a Wishart matrix of the type $\boldsymbol{\mathsf W}=\sfrac{1}{d}\,\boldsymbol{\mathsf X}^T\boldsymbol{\mathsf X}$, where $\boldsymbol{\mathsf X}$ is a $z\times d$ matrix with random Gaussian entries \cite{Parisi2014,Eynard2015,Livan2018}. The probability density of the smaller eigenvalue $\lambda_0$ of $\boldsymbol{\mathsf W}$ scales as $\rho(\lambda)\sim\lambda^{\frac{z-d-1}{2}}$, i.e., in our case ($z=6$, $d=3$), as $\rho(\lambda)\sim\lambda$, that indeed corresponds to a $p_\omega(\theta)\sim\theta^{-3}$ scaling for $\theta\sim\sfrac{1}{\lambda}$. 

\begin{figure} 
\includegraphics[width=\columnwidth]{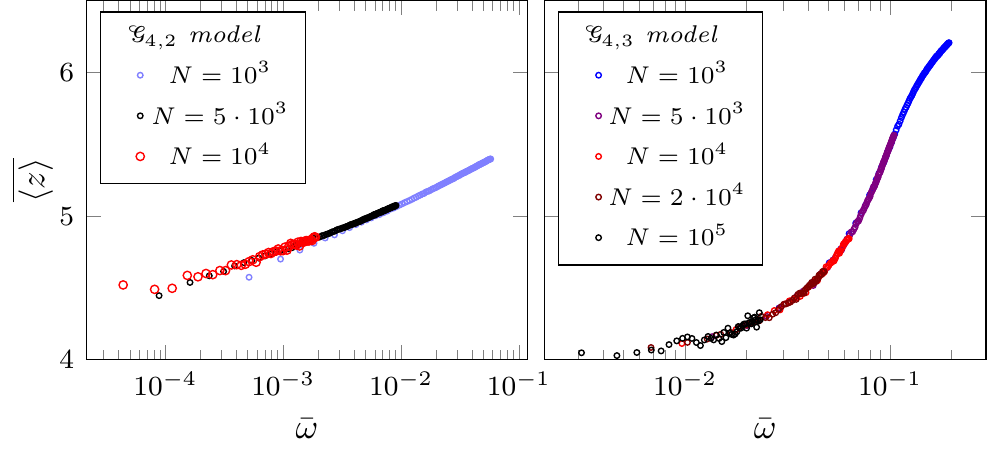}
 \caption{Average coordination $\overline{\langle z\rangle}$ as a function of $\bar \omega$ in the low-frequency region for the $\mG_{4,2}$ model (left) and the $\mG_{4,3}$ model (right), obtained using ED.\label{fig:istoz}}
\end{figure}

A similar behavior is found in the $\mG_{4,2}$ model, but with a different scaling, namely, $p_\omega(\theta)\sim\theta^{-2}$ for large values of $\theta$. This implies, again, a divergent $\mathbb E[\eta_\omega^2]$ for $\omega \to 0$. The different tail scaling in the $\mG_{4,2}$ model can be explained analyzing the average coordination of the $k$th eigenvector $|\boldsymbol{\mathsf{k}}\rangle$,
\begin{equation}\label{zave}
 \overline{\langle z_k\rangle} =\EE{\frac{\sum\limits_{i=1}^N z_i\langle\mathbf k_i|\mathbf k_i\rangle}{\sum\limits_{i=1}^N \langle\mathbf k_i|\mathbf k_i\rangle}}
\end{equation}
that in Fig.~\ref{fig:istoz} we plot as a function of the average frequency $\bar \omega_k\coloneqq\EE{\omega_k}$ of the $k$th eigenvectors. The plot shows that low-frequency modes mostly occupy nodes with low coordination. Assuming that $\langle z\rangle\to 4$ as soon as $\omega \to 0$, the scaling argument proposed for the $\mG_{6,0}$ model can be applied again, and it predicts $p_\omega(\theta)\sim\theta^{-2}$.

The considerations above suggest that in both the $\mG_{6,0}$  and $\mG_{4,2}$ models there is an (avoided) localization transition at $\omega=0$, and the low-frequency modes are extended states that, in the case of the $\mG_{4,2}$ model, have low average coordination.

\subsection{The hyperstatic case}
\begin{figure*}
\subfloat[\label{fig:DOSfixed7all}]{\includegraphics[height=0.58\columnwidth]{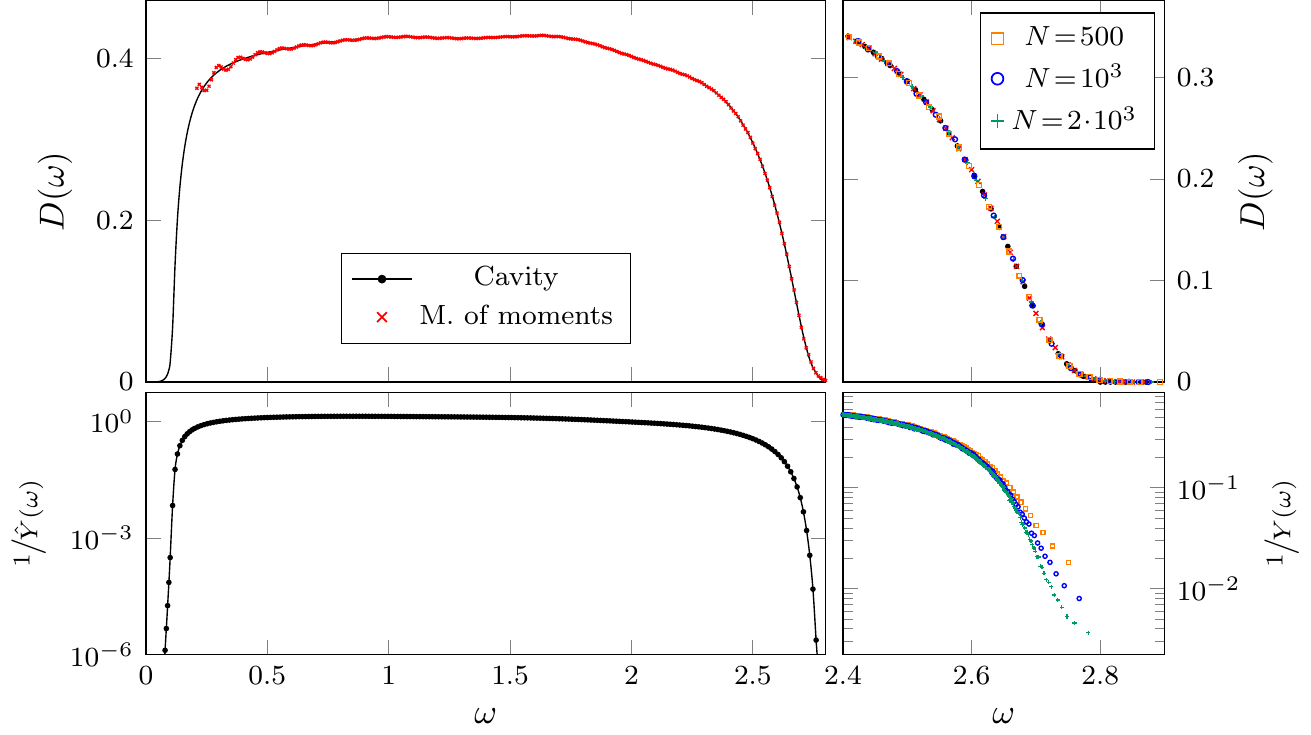}}\hfill
 \subfloat[\label{fig:DOS7}]{\includegraphics[height=0.58\columnwidth]{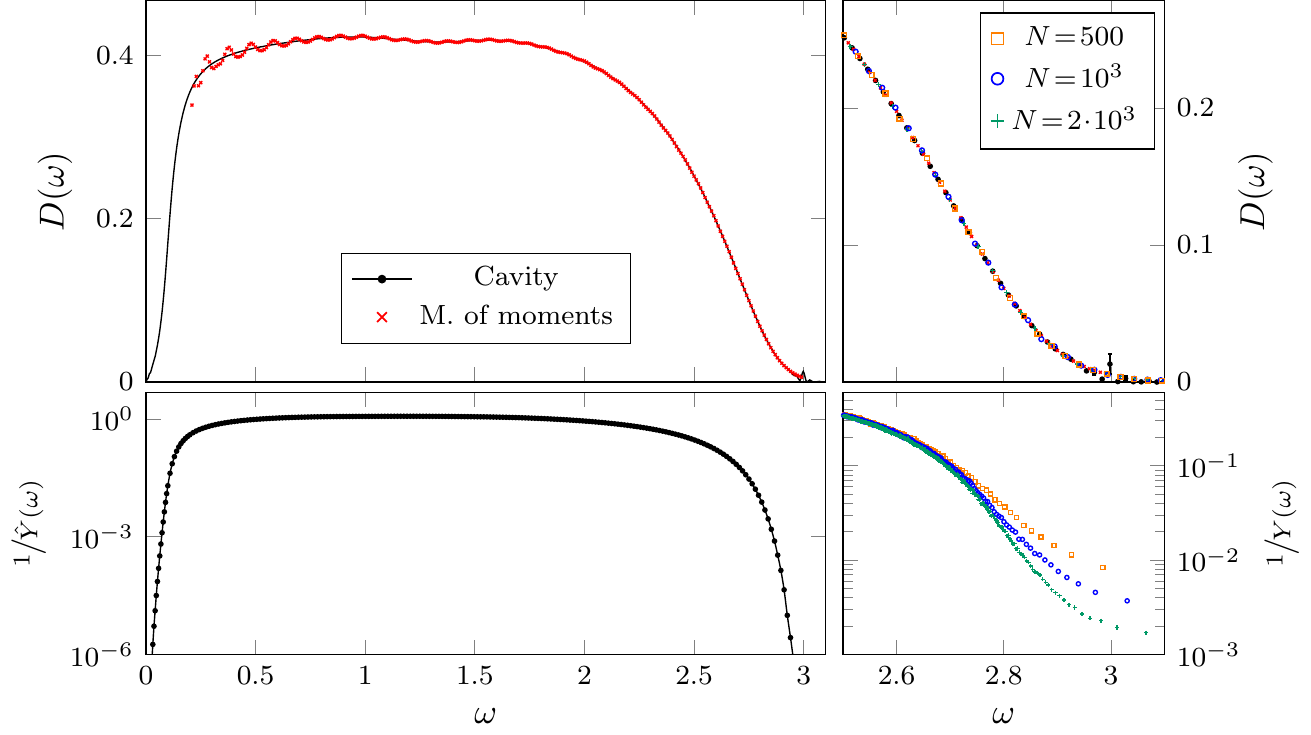}}\hfill
 \caption{DOS, localization indicator $\hat Y$, and participation ratio $\sfrac{1}{Y}$ for the $\mG_{7,0}$ model (a) and the $\mG_{4,3}$ model (b) in the hyperstatic case $\bar z=7$, using the cavity method (black), the method of moments (red), and ED. The numerical integration of the cavity method equation has been performed using $\varepsilon=10^{-8}$ and a population of at least $10^7$ fields for $\omega<0.2$, and a population of $10^6$ fields for the rest of the interval. For the method of moments, $50$ moments were used, averaging over $25$ instances of a matrix $\bM$ with $N=10^6$. The method of moments gives very highly fluctuating results for $\omega<0.2$ that have not been represented. ED results for this region are shown in Fig.~\ref{fig:DOS7head}.}
\end{figure*}

\paragraph{Density of states.} Finally, let us consider the hyperstatic case that, for $d=3$, corresponds to coordination values $\bar z>6$. In this case, a quasi gap opens, and $D(\omega)$ has a power-law behavior for $\omega\to 0$, i.e., $D(\omega)\propto\omega^\alpha$ for some value of $\alpha>0$. As anticipated in the Introduction, the properties that determine the value of $\alpha$ are still a matter of investigation and different results have been found in mean-field models and numerical simulations in finite dimension. Understanding how the finite dimensionality affects the mean-field behavior is of great interest.

\begin{figure*}
\subfloat[\label{fig:DOS7head}]{\includegraphics[height=0.35\textwidth]{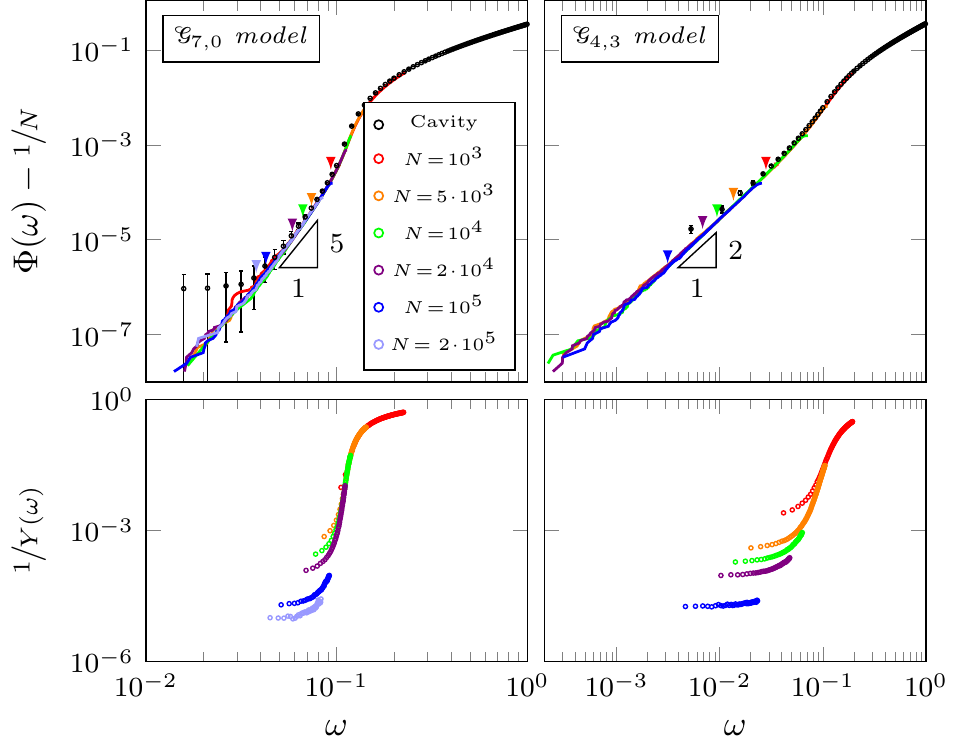}}\hfill
\subfloat[\label{fig:7vari}]{\includegraphics[height=0.35\textwidth]{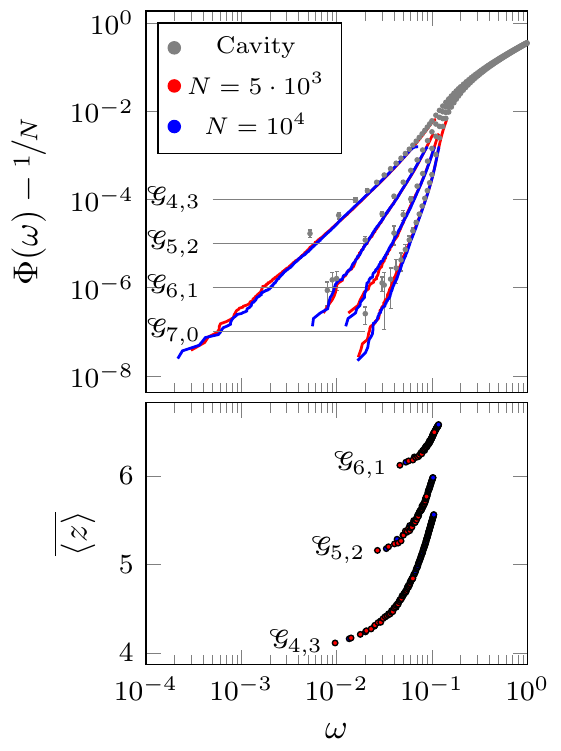}} 
\hfill
\subfloat[\label{fig:omegascaling}]{\includegraphics[height=0.35\textwidth]{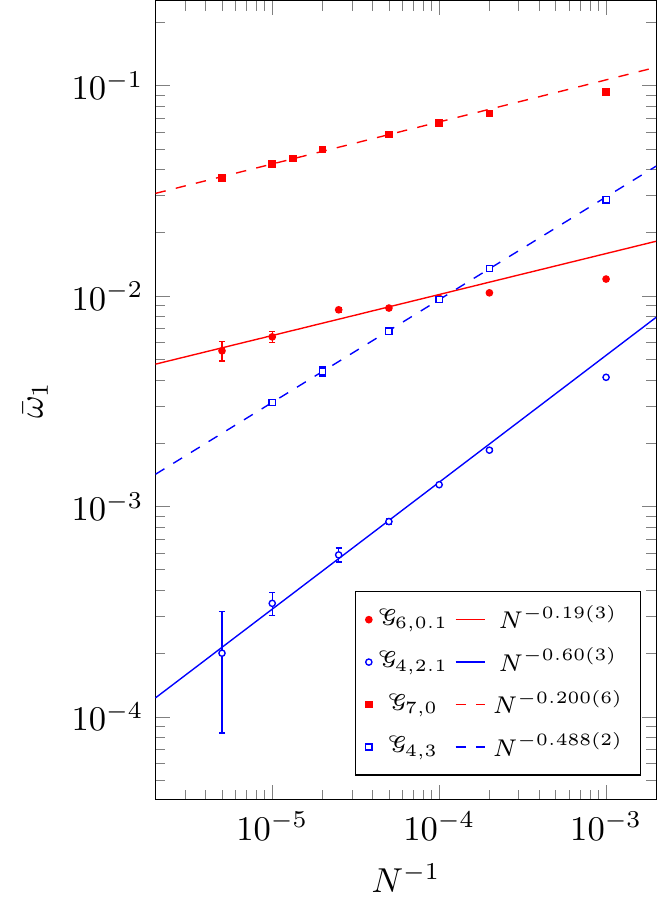}} 
\caption{ED and cavity method results for the hyperstatic regime. (a) Cumulative function (top) and the participation ratio (bottom) at low frequency for the $\mG_{7,0}$ (left) and $\mG_{4,3}$ (right) models. The arrows indicate the value of the average of the first nonzero frequency $\bar\omega_1\coloneqq\EE{\omega_1}$ for each system size. (b) Cumulative function (top) for several systems with $\bar{z}=z_0+\bar\zeta=7$, namely, the $\mG_{7,0}$, $\mG_{6,1}$, $\mG_{5,2}$, and $\mG_{4,3}$ models. On the bottom, the average coordination $\overline{\langle z_k\rangle}$ as a function of the frequency of the $k$th eigenvector [Eq.~\eqref{zave}]. (c) Scaling of the average of the first mode frequency with $N$ in the hyperstatic case. 
The lines are fitted functions.
The fits were performed excluding sizes $N< 10^4$, since for smaller values of $N$ $\bar\omega_1$ is typically located in the bulk and not in the low frequency tail of the distribution (see Figs.~\ref{fig:DOS7head} and \ref{fig:DOS6.1}).
The parameter $a$ of the fit $N^{-a}$ is related to the exponent $\alpha$ by $a=(\alpha+1)^{-1}$ [Eq. \eqref{eq:minScale}], giving
$\alpha=4.3(8)$, $0.67(8)$, $4.0(2)$, and $1.049(8)$ for $\mG_{6,0.1}$, $\mG_{4,2.1}$, $\mG_{7,0}$, and $\mG_{4,3}$, respectively.}
\end{figure*}

The results for the DOS in the $\mG_{7,0}$ model are shown in Fig.~\ref{fig:DOSfixed7all}. Once again, an excellent agreement between the theoretical prediction of the cavity method and the method of moments in the bulk of the spectrum is found. The low-frequency regime is numerically more difficult to evaluate: large system sizes are needed to approach zero frequency with ED. Furthermore, the cavity method itself intrinsically presents some limitations in resolution, due to the finite population in the population dynamics algorithm and the finite value of $\varepsilon$ in the numerical integration of Eqs.~\eqref{ricorsive}. Nevertheless, from the results in Fig.~\ref{fig:DOS7head} we can still find that, for $\omega<10^{-1}$, approximately $\Phi(\omega)\propto\omega^5$ and therefore $D(\omega)\propto\omega^4$, a result that is compatible with theoretical predictions and numerical evidences for finite-dimensional disordered systems \cite{Gurarie2003,Gurevich2003,BaityJesi2015,Lerner2016} and spin glasses on sparse graphs \cite{Lupo2017}. Apart from fitting the low-frequency behavior of $\Phi(\omega)$, the exponent $\alpha$ can be also extracted from the scaling with $N$ of the lowest eigenvalue of the spectrum. Indeed, given a power-law behavior $D(\omega)\sim\omega^\alpha$ for the DOS near the origin, and denoting by $\bar\omega_1\coloneqq\EE{\omega_1}$ the average value of the first mode frequency, we have that
\begin{equation}\label{eq:minScale}
 \int_0^{\bar\omega_1}D(\omega)\dd\omega\sim\frac{1}{Nd}\Rightarrow \bar\omega_1\sim \frac{1}{N^{\frac{1}{\alpha +1}}}.
\end{equation}
This relation has been verified on our data, as shown in Fig.~\ref{fig:omegascaling}, and we find $\alpha=4.0(2)$.

\begin{figure}
\includegraphics[width=0.9\columnwidth]{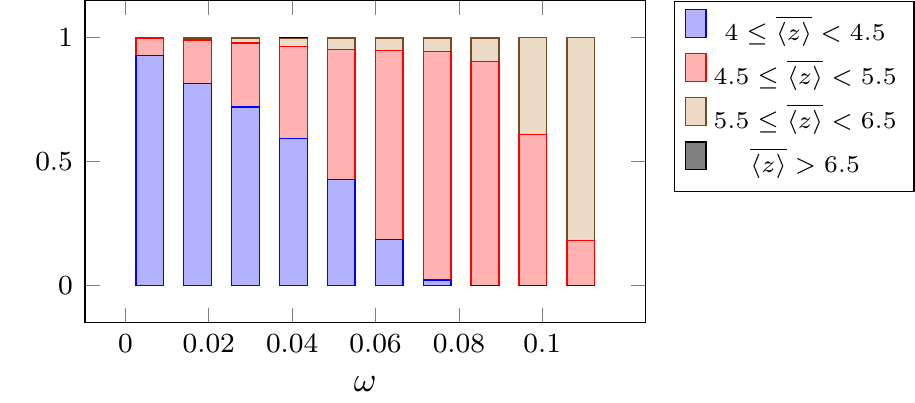}
\caption{Fraction of eigenvalues with different $\overline{\langle z\rangle}$ as a function of $\omega$ in the $\mG_{4,3}$ model. Here we consider the system size $N=5000$. \label{fig:istocoordinazione}}
\end{figure}

A power law behavior, in the same regime, is also found for the $\mG_{4,3}$ model. However, the power law exponent, extracted with the same methods discussed above, is different and we find, in this case, $\Phi(\omega)\propto\omega^2$, i.e., $D(\omega)\propto\omega$. This is also confirmed by the scaling of the first eigenvalue with respect to $N$, as shown in Fig.~\ref{fig:omegascaling}, which gives $\alpha=1.049(8) $. The differences in these results suggest that there is a strong dependence on the topological details of the model, and especially on the lowest admissible coordination, despite the fact that $\bar z$ is the same. Indeed, the lowest part of the spectrum is populated by eigenstates having low average coordination. In Fig.~\ref{fig:istoz} we show that for $\omega<0.1$ the average eigenvalue coordination $\overline{\langle z\rangle}$ --- evaluated using the formula in Eq.~\eqref{zave} --- is below $5$, and asymptotically approaches $4$ as $\omega\to 0$ (see also Fig.~\ref{fig:istocoordinazione}).

To stress the role of the lowest accessible coordination in the power-law exponent $\alpha$, we have also considered the $\mG_{5,2}$ model and the $\mG_{6,1}$ model, both having $\bar z=7$ but with different lowest possible coordination, i.e., $5$ and $6$ respectively. In these cases we observe an intermediate value of the exponent $\alpha$ (see Fig.~\ref{fig:7vari}). In Fig.~\ref{fig:7vari} we have also plotted the $\overline{\langle z\rangle}$ as a function of $\omega$, showing that low-frequency modes are characterized by a low average coordination $\overline{\langle z\rangle}$, close to the lowest coordination allowed by the topology of the graph.

\begin{figure}
\includegraphics[width=\columnwidth]{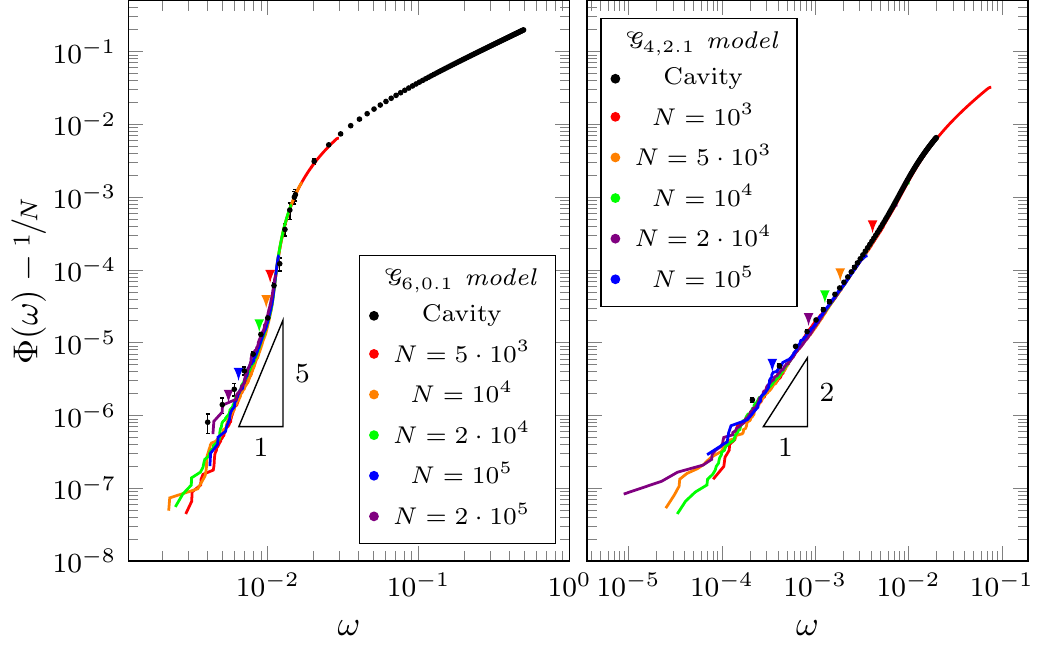}
 \caption{Cumulative function $\Phi(\omega)$ for the $\mG_{6,0.1}$ model and for the $\mG_{4,2.1}$ model using the cavity method (black) and ED (color). The numerical integration of the cavity method equation has been performed using $\varepsilon=10^{-8}$ and a population of $10^6$ fields. The arrows indicate the value of the average of the first nonzero frequency for each system size.\label{fig:DOS6.1}}
\end{figure}

To further exemplify this fact, let us consider a different value of $\bar z$ in the hyperstatic regime: the $\mG_{4,2.1}$ and $\mG_{6,0.1}$ models. Both have the same average coordination $\bar z=6.1$, but they are constructed on a different underlying random regular graph. In the $\mG_{6,0.1}$ model the isostatic condition is realized for every node in the network. Repeating the usual analysis on both models, we obtain the results in Figs.~\ref{fig:omegascaling} and~\ref{fig:DOS6.1}. Similarly to what happens in the $\bar z=7$ case, the results of both the cavity method calculation and ED suggest a different value of $\alpha$ in the two cases: $\alpha=0.67(8)$ for the $\mG_{4,2.1}$ model and $\alpha=4.3(8)$ for the $\mG_{6,0.1}$ model. Note that in the $\mG_{4,2.1}$ model the value of $\alpha$ is closer to the exponent value observed for the $\mG_{4,3}$ model that indeed has the same lowest admissible coordination.

\paragraph{Localization properties.} We present our results on  the localization properties of the eigenstates in the $\mG_{7,0}$ model in Fig.~\ref{fig:DOS7head}. High frequency modes are localized and the IPR scales with the system size for $\omega\gtrsim 2.6$. We also find that there is a low-frequency mobility edge and that for $\omega\lesssim 10^{-1}$  the IPR $Y(\omega)$ scales with the system size. Similarly, localized states are found in the $\mG_{4,3}$ model approximately below the same frequency (see Fig.~\ref{fig:DOS7head}). These results show that, in the hyperstatic regime, at low frequencies a localized region is present. Moreover, taking into account the behavior of $\overline{\langle z\rangle}$ discussed above, in all the analyzed models $\mG_{z_0,\bar\zeta}$ having $\bar\zeta\neq 0$, soft modes appear to be localized on nodes which have very low coordination (see, e.g., Fig.~\ref{fig:soffice}), a fact that is compatible with results of soft sphere systems \cite{Charbonneau2015}. This fact clarifies why the low-frequency behavior of the DOS strongly depends on the lowest possible coordination allowed in the graph topology.

\subsection{Higher dimensions}

\begin{figure}
\includegraphics[width=\columnwidth]{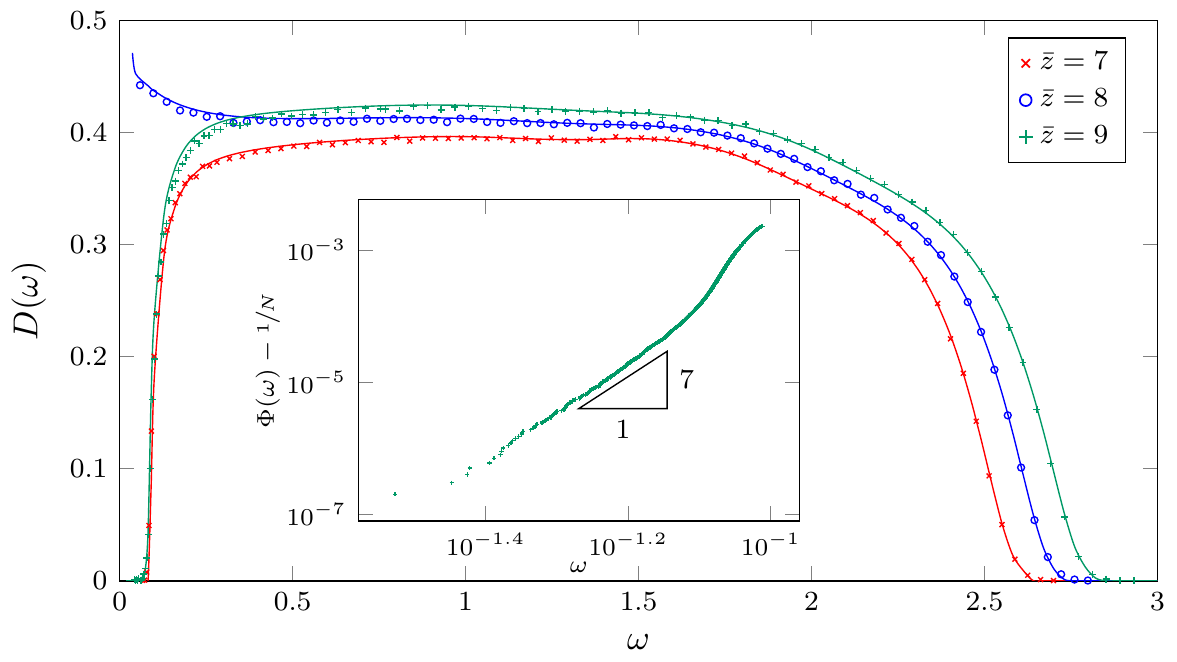}
 \caption{DOS for $d=4$ evaluated on a random regular graph topology for different values of coordination $\bar z$. In the inset, detail of the low-frequency regime. The data have been obtained using ED on the entire spectrum for small values of $N$.  Smooth lines represent the cavity prediction in this case.\label{fig:d4}}
\end{figure}

The behavior of the vibrational DOS in higher dimensions can be studied by changing the dimension of the vector connecting two spheres in contact, $\mathbf{x}_{ij}$ in Eq.~\eqref{Hamhat}. Analyses for $d=4$ on a random regular graph topology show that the DOS follows the expected hypostatic, isostatic, and hyperstatic behavior (see Fig.~\ref{fig:d4}). Specifically, a gap is present in the hypostatic regime, $\bar z=7$, which disappears in the isostatic case $\bar z=2d=8$ and gives way to the expected plateau. In the hyperstatic regime, $\bar z=9$, there is a quasi-gap, and the density of states exhibits a power-law $D(\omega)\propto \omega^6$ (see inset of Fig.~\ref{fig:d4}).

\section{Discussion and conclusions}\label{sec:conclusioni}
In the present work we have discussed a mean-field model for the isostatic transition of soft spheres. The model merges mean-field properties (a contact network defined on a random graph) with finite dimensionality (each contact is associated to a $d$-dimensional vector). We have correctly recovered the main features of the physical isostatic transition, namely, the fact that the average coordination of the graph $\bar z$ determines the general properties of the density of states of the system, $\bar z=2d$ being the isostatic point. 

If $\bar z<2d$ we find a gap in the DOS, and we have verified the scaling of its width with the distance from the isostatic point. For $\bar z\to 2d$ the gap closes.

For $\bar z>2d$ a quasi gap opens. With respect to other mean-field models, such as the perceptron, the model introduced here is able to reproduce additional features that are deeply related to finite-dimensional effects. For example, a localized region is observed at low frequencies. Furthermore, the modes in this region have average coordination typically very close to the lowest possible coordination allowed in the graph, i.e., they are localized on weakly connected nodes. 

The model has enabled us to study the power-law behavior of the DOS $D(\omega)\sim\omega^\alpha$ for $\omega\to 0$ in the hyperstatic regime, in the absence of Goldstone modes. Using both exact diagonalization techniques and the cavity method, we have observed that the exponent $\alpha$ strongly depends on the details of the coordination distribution of the underlying contact network. In particular, the power-law behavior is determined by the aforementioned localized modes and therefore by the lowest accessible coordination in the graph, and not by the average value $\bar z$. Indeed, different models with the same average coordination $\bar z$ but different minimum admissible coordination show different power-law behaviors near the origin. The effect of the finite dimensionality on $\alpha$, and therefore of the finite connectivity, is relevant. {It is, however, worth mentioning that in our model the initial stress contribution has been neglected. It has been very recently observed that this term might be crucial to obtain a $D(\omega)\sim\omega^4$ behavior in the overjammed phase \cite{Lerner2018}. In this sense, the fact that no universal exponent is found in our model might be related to the absence of this contribution.}

This model is an attempt to go beyond the infinite-dimensional models for sphere packings. In the spirit of previous contributions \cite{Parisi2002a,*Grigera2003,Franz2015,*Franz2016,Manning2015}, it relates the spectral properties of disordered systems to a random matrix theory model, combining it with an underlying random graph topology. Moreover, it exemplifies the coordination effects in mean-field models with respect to the spectral properties of amorphous solids \cite{DeGiuli2014}. A large number of open problems remain, such as the precise relation between $\alpha$ and the coordination distribution in the contact network, and further investigations are needed to fill the gap between the finite-dimensional packing problem and the available mean-field models.

\section*{Acknowledgments}
The authors are grateful to F.~Ricci-Tersenghi and  F.~Zamponi for useful discussions. The authors also thank M.L.~Manning, P.~Morse, and E.~Stanifer for discussions and correspondence. F.B, G.P.~and G.S.~acknowledge the financial support of the Simons Foundation (Grant No. 454949, Giorgio Parisi). This work benefited from access to the University of Oregon high-performance computer, Talapas, as well as the Chimera Group cluster at Sapienza Universit\`a di Roma. The work presented in this paper was supported by the project ``Meccanica statistica e complessit\`a'', a research grant funded by PRIN 2015 (Agreement No.~2015K7KK8L). This project has received funding from the European Research Council  (ERC)  under  the  European  Union's  Horizon  2020  research  and  innovation programme (Grant Agreement No.~694925) and benefited from the support of the project THERMOLOC ANR-16-CE30-0023-02 of the French National Research Agency (ANR). 
\appendix

\section{Derivation of the cavity equations}\label{app:ricorsiva}
To derive Eq.~\eqref{ricorsiva} on a sparse graph, let us follow the approach of Refs.~\onlinecite{AbouChacra1973,Cizeau1994}. We consider a generic matrix $\bM$ of size $Nd\times Nd$, such that its element $\mathbf M_{ij}$ is a $d\times d$ submatrix. Pictorially, we can associate the matrix $\bM$ to a graph, in such a way that each Latin index corresponds to a node of the graph, and the submatrix $\mathbf M_{ij}$ is associated to the link $(i,j)$. We also assume that the coordination distribution of the graph is $p_k$. Assuming that $\bM$ is an element of a given ensemble, we are interested in the average DOS of $\bM$ with respect to this ensemble in the $N\to+\infty$ limit. 

It is useful to consider a matrix obtained from $\bM$ creating a ``cavity'' in the graph, i.e., removing a node and/or a link. 
Let us start from the graph corresponding to $\bM$ and let us select, uniformly at random, one of its edges. We then select one of the endpoints of this edge, also at random. This is the node that will be removed. It is called the \textit{cavity node}, and we label it by $0$. We say that the site $0$ is connected to the site $i$ if $\mathbf M_{0i}\neq\mathbf 0$ and/or $\mathbf M_{i0}\neq\mathbf 0$.  It has coordination $\eta_0$, which is distributed as 
\begin{equation}\label{disteta}
\hat p_\eta=\frac{\eta p_\eta}{\sum_{k=1}^\infty kp_k}.
\end{equation}
Observe that if $p_k=\delta_{k,z}$, then $\hat p_\eta=p_\eta=\delta_{\eta,z}$. If instead the coordination follows a Poisson distribution with mean $\lambda$, $p_k=\frac{\lambda^k}{k!}\e^{-\lambda}$, then $\hat p_\eta=\frac{\lambda^{\eta-1}}{(\eta-1)!}\e^{-\lambda}$ with $\eta\geq 1$, i.e., $\sum_\eta \eta \hat p_{\eta}=\lambda+1$.

The cavity graph is simply the graph without the node $0$. 
Once the node is removed, its $\eta_0$ neighbors will have coordination $\eta_i-1$, $i=1,\dots, \eta_0$, where $\eta_i$ are random variables distributed again as in Eq.~\eqref{disteta}. This will be essential for writing down
recursive equations. The matrix $\bM^c$ of the new graph has size $(N-1)d\times(N-1)d$. To proceed in full generality, we will also assume that the removal of the site affects the value of $\mathbf M_{ij}\to \mathbf M_{ij}^c$ for $i,j\neq 0$, due to some required properties of the global matrix that must be preserved, and so the new matrix is not simply a submatrix of the old one with $d$ rows and $d$ columns removed. 

The cavity graph is useful due to the fact that we can find an equation for the elements $\mathbf G_{kk}$ with $k\in\partial 0$ of the cavity resolvent,
\begin{equation}
 \bG(\lambda)\coloneqq \frac{1}{\lambda\bI_{(N-1)d}-\bM^c}
\end{equation}
to be solved in probability. 

Let us now assume that site $0$ is re-introduced but connected to only $\eta_0-1$ of its neighbors \footnote{Note that if the coordination follows a Poisson distribution with mean $\lambda$, the random variable $\eta_0-1$ follows exactly the same distribution.}. Then, $d$ new rows and $d$ new columns are added to the matrix $\bM^c$, obtaining a new matrix $\bM^+$ that has the same dimension of the original matrix but still a ``cavity'', i.e., a missing link. As before, the addition of a site affects in general the entire matrix.  The coordination distribution of the site $0$ is now the same that its neighbors had before its insertion. The new resolvent can be calculated as
\begin{widetext}
\begin{equation}
\begin{split}
\frac{1}{{\left[\bG^{+}\right]}^{\alpha\beta}_{00}(\lambda)}&=\frac{\left[\prod\limits_{k=0}^N\int\dd^d\varphi_k\right] \exp\left(\!-\frac{1}{2}\sum\limits_{{k,l=0}}^N\sum\limits_{{\mu,\nu=1}}^d  \varphi_k^\mu\left[\lambda\bI_{(N+1)d}-\bM^+\right]_{kl}^{\mu\nu}\varphi_l^\nu\right)}{\left[\prod\limits_{k=0}^N\int\dd^d\varphi_k\right] \varphi_0^\alpha \varphi_0^\beta\exp\!\left(\!-\frac{1}{2}\sum\limits_{{k,l=0}}^N\sum\limits_{{\mu,\nu=1}}^d  \varphi_k^\mu\left[\lambda\bI_{(N+1)d}-\bM^+\right]_{kl}^{\mu\nu}\varphi_l^\nu\right)}\\
 &=\left[\lambda\bI_{d}-\mathbf M_{00}^+-\sum_{k,l\neq 0}\mathbf M^+_{0k}\cdot \frac{1}{\lambda\delta_{kl}\mathbf I_d-\mathbf M^+_{kl}}\cdot\mathbf M^+_{l0}\right]^{\alpha\beta}.
\end{split}
\end{equation}
\end{widetext}
Remembering now that, for $k,l\neq 0$
\begin{equation}
 \lambda\bI_d-\mathbf M^+_{kl}=\left[\frac{1}{\bG(\lambda)}\right]_{kl}-\left(\mathbf M^+_{kl}-\mathbf M_{kl}^c\right),
\end{equation}
and denoting by $\boldsymbol\Delta_{kl}\coloneqq \mathbf M^+_{kl}-\mathbf M_{kl}^c$ we can write \cite{Ciliberti2003}
\begin{multline}
 \frac{1}{{\left[\bG^{+}\right]}^{\alpha\beta}_{00}}=\\
 =\left[\lambda\bI_{d}-\mathbf M^+_{00}-\sum_{k,l\neq 0}\mathbf M^+_{0k}\cdot \frac{1}{\left[\frac{1}{\bG}\right]_{kl}-\boldsymbol\Delta_{kl}}\cdot\mathbf M^+_{l0}\right]^{\alpha\beta}.
\end{multline}

Let us now specify the equations above to our problem. In the case of a symmetric dynamical matrix in the form in Eq.~\eqref{proprieta}, due to the rule in Eq.~\eqref{proprieta}, for $i,j\neq 0$, $\boldsymbol\Delta_{ij}=-\delta_{ij}\mathbf M^+_{i0}$. Using the fact that $\mathbf M_{00}^+=-\sum_{k\in\partial 0}\mathbf M_{k0}^+$, the recursive equation becomes
\begin{multline}
 \frac{1}{{\left[\bG^{+}\right]}^{\alpha\beta}_{00}}=\\
 =\left[\lambda\bI_{d}+\sum_{\mathclap{k\in\partial 0}}\mathbf M_{k0}^+-\sum_{\mathclap{k,l\in\partial 0}}\mathbf M^+_{0k}\!\cdot\!\frac{1}{\left[\frac{1}{\bG}\right]_{kl}\!\!+\delta _{kl}\mathbf M_{k0}^+}\!\cdot\!\mathbf M_{l0}^+\right]^{\alpha\beta}.
\end{multline}
The sums in the equation above run over the $\eta-1$ neighbors of the vertex $0$. In the case of a sparse random graph we have that any two neighbors of $0$, let us say $k$ and $l$, are almost surely \textit{not} directly connected for $N\to+\infty$ and therefore
\begin{equation}
 \left[\frac{1}{\bG}\right]_{kl}=-\mathbf M^+_{kl}\equiv \boldsymbol 0.
\end{equation}
Moreover, if we assume that the off diagonal submatrices $\mathbf G_{ij}$ are subleading for $i\neq j$,
\begin{equation}
\left[\frac{1}{\bG}\right]_{kk}=\frac{1}{\mathbf G_{kk}-\sum_{l\in\partial k}\mathbf G_{kl}\cdot\left[\frac{1}{\bG}\right]_{ll}\cdot \mathbf G_{lk}}\approx\frac{1}{\mathbf G_{kk}}.
\end{equation}
Using this observation, and the fact that $\mathbf M^+_{0k}$ is a projector, Eq.~\eqref{ricorsiva} can be obtained from
\begin{multline}
 \sum_{k\in\partial 0}\mathbf M_{k0}^+-\sum_{k,l\in\partial 0}\mathbf M^+_{0k}\cdot\frac{1}{\left[\frac{1}{\bG}\right]_{kl}+\delta _{kl}\mathbf M_{k0}^+}\cdot\mathbf M_{l0}^+\\
 =\sum_{k\in\partial 0}\sum_{n=0}^{\infty}\left(-\mathbf M_{0k}^+\cdot \frac{1}{\left[\frac{1}{\bG}\right]_{kk}}\right)^n\cdot\mathbf M_{k0}^+\\
 =\sum_{k\in\partial 0}\frac{\mathbf M^+_{0k}}{1+\tr\left(\frac{1}{\left[\frac{1}{\bG}\right]_{kk}}\cdot \mathbf M_{0k}^+\right)}\\
 \approx \sum_{k\in\partial 0}\frac{\mathbf M^+_{0k}}{1+\tr\left(\mathbf G_{kk}\cdot \mathbf M_{0k}^+\right)}.
\end{multline}
Observe that the right-hand side of the previous equation depends only on the elements of $\bG$ corresponding to the neighbors of the cavity site $0$ before its insertion. Due to the randomness in the model, it is not true in general that a fixed point solution of Eq.~\eqref{ricorsiva} exists. However, we expect that the equation is true in probability, and we can search for a fixed point in the space of probability distributions of $\mathbf G$, solving the equation by means of a population dynamics algorithm. 
The fixed-point population of $\mathbf G$ that is found corresponds to a resolvent evaluated on a node of the graph with $\eta-1$ neighbors. The ``true'' local resolvent $\mathbf R$ for a site with $z$ neighbors distributed with probability $p_z$ can be obtained performing one last step, given by Eq.~\eqref{ricorsiva2}, extracting the $z$ required elements $\mathbf G_k$ from the cavity field population.
\section{The method of moments}\label{app:mom}
In this Appendix, we summarize the method of moments that we used to compute the DOS of the Hessian matrix in Eq.~\eqref{proprieta}. We will give here the procedure only, without providing the necessary proofs that can be found in the literature\cite{Cyrot-Lackmann1967,*Gaspard1973,*Lambin1982,*Jurczek1985, *Benoit1992,*Villani1995}. The method, as opposed to ED, does not determine the single eigenvalues if the number of moments used are less than the rank of the Hessian matrix. Instead, it gives the envelope of their density. This has the advantage of allowing access to the entire spectrum even when using a limited number of moments.

Let us start by assuming that an $N\times N$ matrix $\bM$ is given and that we want to evaluate a spectral density function of the form
\begin{equation}
\phi_{\bp}(\lambda)\coloneqq\sum_{k=1}^N |\langle \bp|\bk\rangle|^2\delta(\lambda-\lambda_k).
\end{equation}
In the equation above, $\lambda_k$ is the $k$th eigenvalue of the matrix $\bM$ with corresponding eigenvectors $|\bk\rangle$, $\bM|\bk\rangle=\lambda_k|\bk\rangle$, and $|\bp\rangle$ is a given vector. If we introduce the Stiltjes transform
\begin{equation}
R(z)\coloneqq \int_{-\infty}^{\infty}\frac{\phi_\bp(\lambda)}{z-\lambda}\dd\lambda,
\end{equation}
then the following relation holds:
\begin{equation}\label{rphimom}
\phi_\bp(\lambda)=-\frac{1}{\pi}\lim_{\varepsilon\to 0}\Imm R(\lambda+i\varepsilon).
\end{equation}

The non-negative function $\phi_\bp(\lambda)$ can be used as a weight function to generate a sequence of orthogonal polynomials $p_{n}(z)$ by imposing
\begin{equation}
\int \lambda^np_n(\lambda)\phi_\bp(\lambda)\dd\lambda=0.
\end{equation}
These polynomials satisfy the relation
\begin{subequations}
\begin{align}
p_{-1}(\lambda)&=0,\\
p_{0}(\lambda)&=1,\\
p_n(\lambda)&=(\lambda-a_n)p_{n-1}(\lambda)-b_{n-1}p_{n-2}(\lambda),\  n=1,2,\dots\label{ricorsivamom}
\end{align}\end{subequations}
where
\begin{eqnarray}
a_{n}\coloneqq\frac{\bar \nu_{n-1}}{\nu_{n-1}},\quad b_{n}\coloneqq\frac{\nu_{n}}{\nu_{n-1}},
\end{eqnarray}
and 
\begin{equation}
\nu_n\coloneqq \int p_n^2(\lambda)\phi_\bp(\lambda)\dd\lambda,\quad \bar\nu_n\coloneqq \int \lambda p_n^2(\lambda)\phi_\bp(\lambda)\dd\lambda
\end{equation}
are the generalized moments of $\phi_\bp(\lambda)$. 

The method relies on the nontrivial fact that the coefficients $a_n$ and $b_n$ in the recurrence relation for the polynomials $p_n(\lambda)$ are the same as those in the representation of $R(z)$ as a continued Jacobi fraction, i.e.,
\begin{equation}
 R(z) = \cfrac{1}{z-a_1-\cfrac{b_1}{z-a_2-\cfrac{b_2}{z-a_3+\dots}}}.
\end{equation}
This implies that, truncating the continued fraction expansion for $R$ to some order $M$, we can estimate $\phi_\bp$ by means of a finite set of coefficients $\{a_n,b_n\}$, i.e., a finite set of generalized moments. Moreover, it turns out that the generalized moments can be evaluated very easily by a sequence of matrix multiplications. Starting from the normalized vector
\begin{equation}
|\bt_0\rangle\coloneqq\frac{1}{\sqrt{\langle\bp|\bp\rangle}}|\bp\rangle,
\end{equation}
we can apply to it the recursive relation
\begin{equation}
|\bt_{n+1}\rangle=\left(\bM-a_{n+1}\bI_N\right)|\bt_n\rangle-b_n|\bt_{n-1}\rangle,
\end{equation}
and extract the coefficients using
\begin{equation}
\nu_n=\langle\bt_n|\bt_n\rangle,\quad \bar\nu_n=\langle\bt_n|\bM|\bt_n\rangle.
\end{equation}
By evaluating $\{a_n\}_{n=1,\dots M}$ and $\{b_n\}_{n=1,\dots M}$ up to a certain order $M$ we can finally reconstruct $R(z)$ and then $\phi_\bp(\lambda)$ by means of Eq.~\eqref{rphimom}. The spectral density 
\begin{equation}
\rho(\lambda)=\frac{1}{N}\sum_{i=1}^N\delta(\lambda-\lambda_i)
\end{equation}
can be obtained averaging $\phi_\bp(\lambda)$ over all possible vectors $|\bp\rangle$, being $\overline{|\langle\bp|\bk\rangle|^2}=\sfrac{1}{N}$.

When a high number of moments is used ($M\approx 100$) numerical stability is further improved by performing a Gram-Schmidt orthonormalization of the vectors $|\bt_n\rangle$ at every iteration step. Finally, a truncation term $T(z)$ can be added to take into account the neglected terms in the continued fraction, i.e.,
\begin{equation}
 R(z) = \cfrac{1}{z-a_1-\cfrac{b_1}{z-a_2\dots-\cfrac{b_n}{z-a_n+T(z)}}}.
\end{equation}
Assuming that $a_n\to a$ and $b_n\to b$ when $n\to\infty$, with at most small oscillations around these values, $T(z)$ can be estimated from
\begin{equation}
 T(z)=\frac{1}{z-a-bT(z)}.
\end{equation}
For details on the stability and precision of the method, we refer to Refs.~\onlinecite{Cyrot-Lackmann1967,*Gaspard1973,*Lambin1982,*Jurczek1985, *Benoit1992,*Villani1995}.

\bibliography{biblio.bib}
\end{document}